\documentclass[journal,draftclsnofoot,onecolumn,12pt,twoside]{IEEEtranTCOM}
\usepackage{graphicx}
\usepackage{caption}
\usepackage{subcaption}
\usepackage{float}
\usepackage{cite}
\usepackage{amsfonts}
\usepackage{multicol}
\usepackage{amsmath}

\newtheorem{lemma}{Lemma}
\newtheorem{proposition}{Proposition}

\normalsize

\begin{document}

\title{On the Performance of X-Duplex Relaying}

\author{\IEEEauthorblockN{Shuai Li\IEEEauthorrefmark{1}, Mingxin Zhou\IEEEauthorrefmark{1}, Jianjun Wu\IEEEauthorrefmark{1}, Lingyang Song\IEEEauthorrefmark{1}, \\Yonghui Li\IEEEauthorrefmark{2}, and Hongbin Li\IEEEauthorrefmark{1}} \IEEEauthorblockA{\\\IEEEauthorrefmark{1} School of Electronics Engineering and Computer Science, \vspace{-2mm}\\Peking University, Beijing, China \vspace{-2mm}\\(E-mail:
{shuai.li.victor, mingxin.zhou, just, lingyang.song, lihb}@pku.edu.cn)\\ \IEEEauthorrefmark{2}School of Electrical and Information Engineering, \vspace{-2mm}\\The University of Sydney, Australia \vspace{-2mm}\\(E-mail: yonghui.li@sydney.edu.au)\\  }
\thanks{This work has been partially accepted by IEEE ICC 2016~\cite{shuai}.}
}

\maketitle
 \thispagestyle{empty}
\pagestyle{empty}
\vspace{-20mm}
\begin{abstract}
%\boldmath
In this paper, we study a X-duplex relay system with one source, one amplify-and-forward (AF) relay
and one destination, where the relay is equipped with a shared antenna and two radio frequency (RF) chains used for transmission or reception.
X-duplex relay can adaptively configure the connection between its RF chains and antenna to operate in either HD or FD mode, according to the instantaneous channel conditions.
%We analyze the system overall performance of X-duplex relaying based on
We first derive the distribution of the signal to interference plus noise ratio (SINR), based on which we then analyze the outage probability, average symbol error rate (SER), and average sum rate.
We also investigate the X-duplex relay with power allocation and derive the lower bound and upper bound of the corresponding outage probability. Both analytical and simulated results show that the X-duplex relay achieves a better performance over pure FD and HD schemes in terms of SER, outage probability and average sum rate, and the performance floor caused by the residual self interference can be eliminated using flexible RF chain configurations.
\end{abstract}

\vspace{-5mm}

\IEEEpeerreviewmaketitle
 \thispagestyle{empty}

\section{Introduction}

\IEEEPARstart{F}{ull-duplex} (FD) enables a node to receive and transmit information over the same frequency simultaneously \cite{choi2010achieving}.
Compared with half-duplex (HD), FD can potentially enhance the system spectral efficiency due to its efficient bandwidth utilization.
However, its performance is affected by the self interference caused by signal leakage in FD radios \cite{Shong}.
The self interference can be suppressed by using digital-domain \cite{duarte2010Fullduplex,duarte2012experiment,day2012full}, analog-domain \cite{Everett2011empower,MJain2011,EAryafar2012} and propagation-domain methods
\cite{riihonen2010residual,riihonen2011mitigation,senaratne2011beamforming}.
However, the residual interference still exists due to imperfect cancellation \cite{everett2014passive,sabharwal2013band}.
% \renewcommand{\thefootnote}{}
%\footnotetext{This work has been partially accepted by IEEE International Conference on Communications (ICC) 2016.}

Recently, FD technique has been deployed into relay networks \cite{HJU,riihonen2009feasibility}.
%Compared with HD, FD relaying is more spectrally efficient \cite{}.
The capacity trade off between FD and HD in a two hop AF relay system is studied \cite{riihonen2009comparison}, where the source-relay and the self interference channels are modeled as non-fading channels.
%Authors in \cite{riihonen2009comparison} studied the capacity trade-off between FD and HD modes in a two-hop AF relay wireless communication system, in which the source-relay and the self interference channels are modeled as non-fading channels.
The two-hop FD decode-and-forward (DF) relay system was analyzed in terms of the outage event, and the conditions that FD relay is better than HD in terms of outage probability were derived in \cite{Taehoon2010optimal}.
The work in \cite{krikidis2012full} analyzed the outage performance of an optimal relay selection scheme with dynamic FD/HD switching based on the global channel state information (CSI).
In \cite{7065323}, the authors analyzed the multiple FD relay networks with joint antenna-relay selection and achieved an additional spatial diversity than the conventional relay selection scheme.

Though FD has the potential to achieve higher spectrum efficiency than HD, HD outperforms FD in the strong self interference region.
%The work in \cite{riihonen2011hybrid} proposed the hybrid FD/HD switching and optimized the instantaneous and average spectral efficiency in a two-antenna infrastructure relay system.
The work in \cite{riihonen2011hybrid} proposed the hybrid FD/HD switching and optimized the instantaneous and average spectral efficiency in a two-antenna infrastructure relay system.
For the instantaneous performance, the optimization is studied in the case of static channels during one instantaneous snapshot within channel coherence time and the distribution of self interference is not considered.
For the average performance, the self interference channel is modelled as static.
%where the channel fading of self interference channel is considered as slow-fading or non-fading in the optimization.
The outage probability and ergodic capacity for two-way FD AF relay channels were investigated while the self interference channels are simplified as additive white Gaussian noise channels in \cite{Rongyi}.
In practical systems, the residual self interference can be modeled as the Rayleigh distribution due to multipath effect\cite{tmk,7065323,krikidis2012full}.
In this case, the analysis becomes a non-trivial task.
%The X-duplex relay scheme, where the relay with one shared anntenna adaptively selects between FD and HD modes based on the instantaneous channel conditions, is proposed in this paper.
%To the best of our knowledge, the cumulative distribution function (CDF) of the equivalent SINR of the X-duplex relay scheme have not been derived. In this paper, we will derive the asymptotic CDF expression of the SINR of the X-duplex relay scheme with a shared antenna relay\cite{bharadia2014full,chen2013flexradio,Wang2014}.

In this paper, we consider a FD relay system consisting of one source node, one AF relay node and one destination node.
Different from existing works on FD relay with predefined RX and TX antennas, in our paper, the relay node is equipped with an adaptively configured shared antenna, which can be configured to operate in either transmission or reception mode \cite{HJU1,HJU2,bharadia2014full,chen2013flexradio,Wang2014}.
The shared antenna deployment can use the antenna resources more efficiently compared with separated antenna as only one antenna set is adopted for both transmission and reception simultaneously \cite{HJU2,Gliu2015}.
One shared-antenna is more suitable to be deployed into small equipments, such as mobile phone, small sensor nodes, which is essentially different from separated antennas in terms of implementation \cite{riihonen2011hybrid}.
The relay can select between FD and HD modes to maximize the sum rate by configuring the relay node with a shared antenna based on the instantaneous channel conditions.
We refer to this kind of relay as a X-duplex relay.
%Different from the hybrid FD/HD switching scheme in \cite{riihonen2011hybrid}, the X-duplex relay is regarded as a concept of smart relay which can determine the optimal mode, schedule the transmission and complete the resource allocation.
%The concept of X-duplex relay treats the relay as a smart relay that composes one whole system, while the hybrid FD/HD mode switching scheme considers two mode as individual systems and maximize the system performance by obtaining benefits from each mode \cite{riihonen2011hybrid}.

First, the asymptotic CDF of the received signal at the destination of the X-duplex relay system is calculated, then, the asymptotic expressions of outage probability, average SER and average sum rate are derived and validated by Monte-Carlo simulations.
We show that the X-duplex relay can achieve a better performance compared with pure FD and HD modes and can completely remove the error floor due to the residual self interference in FD systems.
To further improve the system performance, a X-duplex relay with adaptive power allocation (XD-PA) is investigated where the transmit power of the source and relay can be adjusted to minimize the overall SER subject to the total power constraint.
The end-to-end SINR expression is calculated and a lower bound and a upper bound are provided.
The diversity order of XD-PA is between one and two.

The main contributions of this paper are listed as follows:

1) The X-duplex relay with a shared antenna is investigated in a single relaying network, which can increase the average sum rate.

2) Taking the residual self interference into consideration, the CDF expression of end-to-end SINR of the X-duplex relay system is derived.

3) The asymptotic expressions of outage probability, average SER and average sum rate are derived based on the CDF expression and validated by simulations.

4) Adaptive power allocation is introduced to further enhance the system performance of the X-duplex relay system. A lower bound and an upper bound of the outage probability of XD-PA are derived and the diversity order of XD-PA is analyzed.

The remainder of this paper is organized as follows: In Section \uppercase\expandafter{\romannumeral2}, we introduce the system model and X-duplex relay. In Section \uppercase\expandafter{\romannumeral3}, the outage probability, the average SER and the average sum rate of the X-duplex relay system are derived and a lower bound and a upper bound of the end-to-end SINR of XD-PA are provided.
Simulation results are presented in Section \uppercase\expandafter{\romannumeral4}. We draw the conclusion in Section \uppercase\expandafter{\romannumeral5}.

%\IEEEpubidadjcol
\section{System Model}

As shown in Fig. 1, we consider a system which consists of one source node (S), one destination node (D), and one AF relay node (R). We assume the direct link from S to D is strongly attenuated and information can only be forwarded through the relay node. In this network, all nodes operate in the same frequency and each of them is equipped with one antenna. Node R is equipped with one transmit (TX) and one receive (RX) RF chains which can receive and transmit signal over the same frequency simultaneously\cite{bharadia2014full}. In the X-duplex relay, node R can adaptively switch between the FD and HD modes according to the residual self interference between the two RF chains of the relay node and the instantaneous channel SNRs between the source/destination node and relay node. In this paper, all the links are considered as block Rayleigh fading channels. We assume the channels
remain unchanged in one time slot and vary independently from one slot to another.
The derivation of end-to-end SINR of FD and HD mode is similar to the discussions in the earlier works\cite{riihonen2009feasibility,riihonen2011hybrid}.

%\begin{figure}
%\begin{minipage}[t]{0.5\linewidth}
%\centering
%\includegraphics[width=2.2in]{FD2.eps}
%\caption{The FD relay.}
%\label{Fig.sub.1}
%\end{minipage}%
%\begin{minipage}[t]{0.5\linewidth}
%\centering
%\includegraphics[width=2.2in]{HD.eps}
%\caption{The HD relay.}
%\label{Fig.sub.2}
%\end{minipage}
%\caption{\small System model of X-duplex Relay.}
%        \label{fig:2}
%\end{figure}

\begin{figure}[H]
        \begin{subfigure}[b]{0.5\linewidth}
            \includegraphics[width=\textwidth]{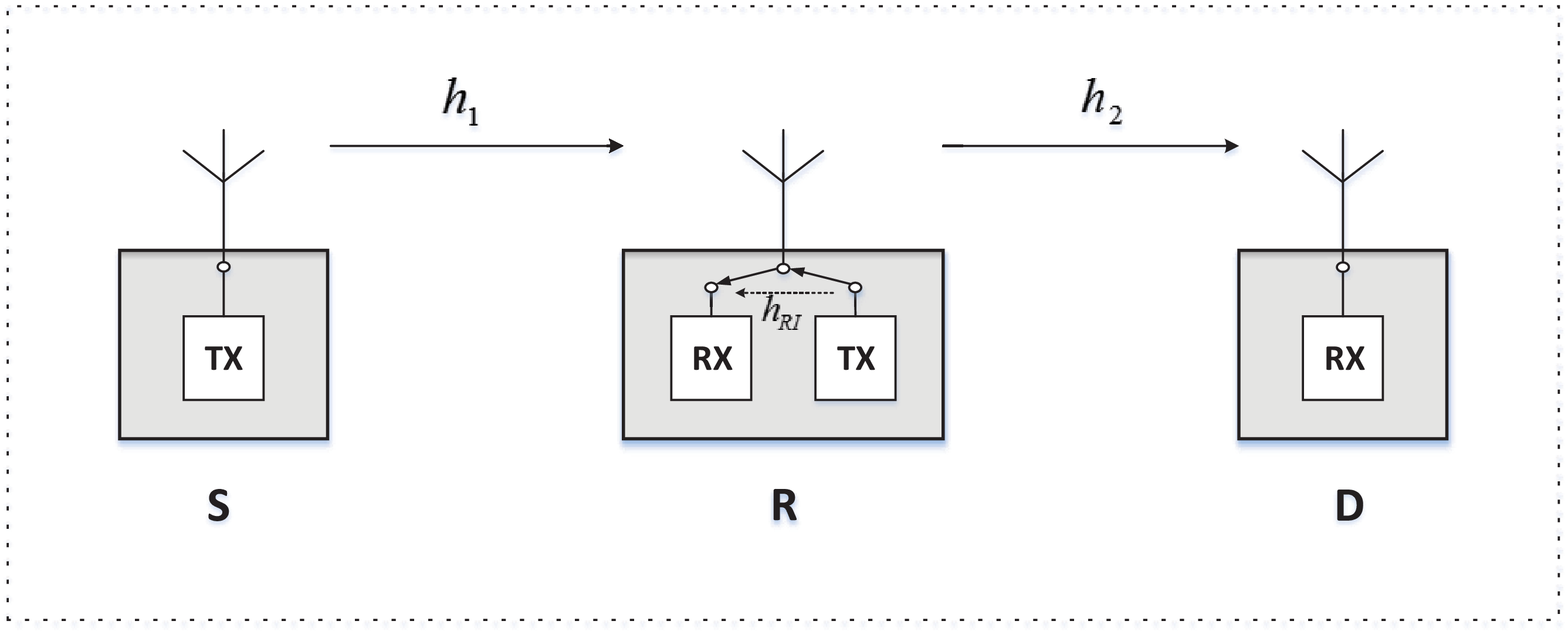}
            \caption{{{The FD relay.}}}
            \label{Fig.sub.1}
        \end{subfigure}
        \quad
        \begin{subfigure}[b]{0.5\linewidth}
            \includegraphics[width=\textwidth]{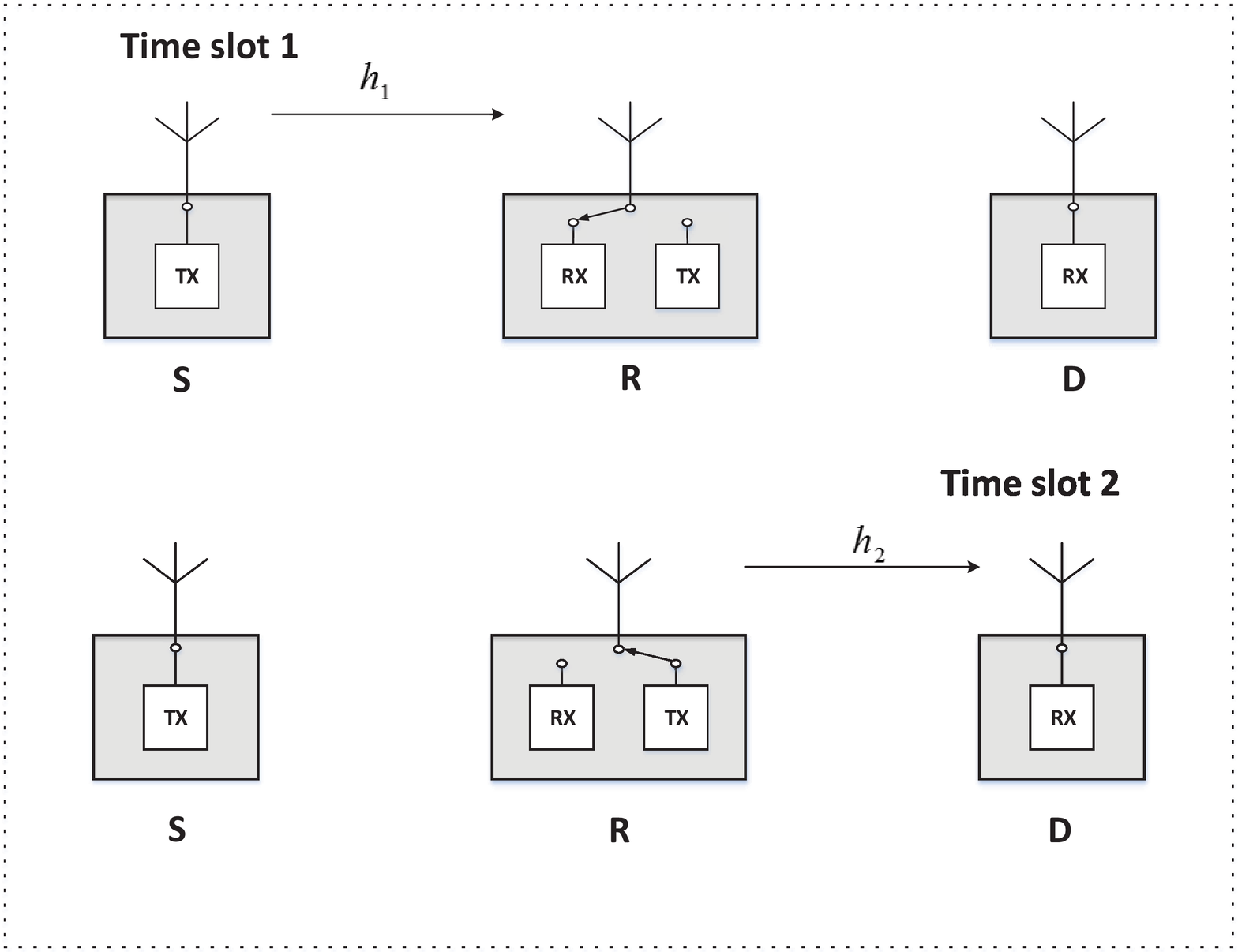}
            \caption{{The HD relay.}}
            \label{Fig.sub.2}
        \end{subfigure}
         \caption{System model of X-duplex Relay.}
        \label{fig:2}
        \vspace{-8mm}
\end{figure}

%\begin{figure}[H]
%
%\centering
%
%\subfigure[The FD relay]{
%
%\label{Fig.sub.1}
%
%\includegraphics[width=4in]{FD2.eps}}
%
%\subfigure[The HD relay]{
%
%\label{Fig.sub.2}
%
%\includegraphics[width=4in]{HD.eps}}
%
%\caption{System model of X-duplex Relay.}
% \vspace{-8mm}
%\label{fig:2}
%\end{figure}
\subsection{End-to-End SINR }
In the FD mode, both RX and TX chains at node R are active at the same time. The signal received at node R is given as
\begin{equation}\label{lisignal}
{y_r} = {h_1}\sqrt {{P_S}} x + {h_{RI}}\sqrt {{P_R}} {x_r} + {n_1},
\end{equation}
where ${h_1}$ denotes the channel between source and relay, $h_{RI}$ is the residual self interference of relay R. $x$ and ${x_r}$ denote the transmit signal of the source and relay. ${P_S}$ and ${P_R}$ are the transmit powers of the source and relay node. ${n_1}$ is the zero-mean-value additive white Gaussian noise with the power ${\sigma ^2}$.

%The desired receiving signal at the relay in (1) is ${h_1}\sqrt {{P_S}} x$  and  self-interference cancellation is adopted. to mitigate ${h_{RI}}\sqrt {{P_R}} {x_r}$.
%However, due to the non-ideal CSI, insufficient dynamic range and other factors in the signal process, ideal SIC can not be realized.
%Especially, the source signal might behave as interference to the SIC schemes in the active cancellation.

%We denote the residual noise with $\sqrt {{P_R}} \widehat {{h_{RI}}}\widehat {{x_r}}$.
%Therefore, (1) is transformed into ${y_r} = {h_1}\sqrt {{P_S}} x + \widehat {{h_{RI}}}\sqrt {{P_R}} \widehat {{x_r}} + {n_1}$, where $E[|\widehat {{x_r}}{|^2}] \le 1$ and $\widehat {{h_{RI}}}$ represents the RSI due to imperfect SIC.
%As cancellation algorithms [4]-[12] are mainly characterized by specific residual
%power, parametrization by ${\left| {\widehat {{h_{RI}}}} \right|^2}$ makes the analysis more
%generic.
%If only passive suppression is adopted, $\widehat {{h_{RI}}}\widehat {{x_r}} = {h_{RI}}{x_r}$.

%In the revised paper, $\eta$ is given as $\eta  = \frac{{{\lambda _R}{P_R}}}{{{\lambda _1}{P_S}}}$, where $P_S$, $P_R$ are the transmit power of the source and relay, $\lambda_1$ is the mean value of instantaneous SNR $\gamma_1$, $\lambda_R$ is the mean value of RSI gain ${\gamma _R} = {\left| {\widehat {{h_{RI}}}} \right|^2}/{\sigma ^2}$ due to imperfect SIC.
%Also, the parameter $\eta$ is not considered as a fixed ratio value.
%Basically, $\eta$ is adopted to simplify the derivation in our analysis.

AF protocol is adopted at relay R and the forwarding signal at the relay R can be written as
\begin{equation}
{x_r} = {\beta _f} \cdot {y_r},
\end{equation}
where ${\beta _f}$ denotes the power amplification factor satisfying
%Factor ${\beta _f}$ ensures the average power of ${x_r}$ to fulfill the power constraint at relay R.
\begin{equation}
E[| {x_r}{|^2}] = {\beta _f}^2(|{h_1}{|^2}{P_S} + | {h_{RI}}{|^2}{P_R} + {\sigma ^2}) \le 1,
\end{equation}
where
\begin{equation}\label{betaf}
{\beta _f}^2 = \frac{1}{{|{h_1}{|^2}{P_S} + | {h_{RI}}{|^2}{P_R} + {\sigma ^2}}}.
\end{equation}

The received signal at the destination D is given by
\begin{equation}
{y_d} = {h_2}\sqrt {{P_R}} {x_r} + {n_2},
\end{equation}
where ${h_2}$ denotes the channel between relay and destination, and ${n_2}$ is the zero-mean-value additive white Gaussian noise with power ${\sigma ^2}$.

The end-to-end SINR of FD mode can be expressed as
\begin{equation}
{\gamma _F} = \frac{{{P_S}{P_R}|{h_1}{|^2}|{h_2}{|^2}{\beta _f}^2}}{{{P_R}^2|{h_2}{|^2}| {h_{RI}}{|^2}{\beta _f}^2 + {P_R}|{h_2}{|^2}{\beta _f}^2{\sigma ^2} + {\sigma ^2}}},
\end{equation}
using (\ref{betaf}) , the SINR can be further simplified as
\begin{equation}\label{fdsinr}
{\gamma _F} = \frac{{{P_S}{P_R}{\gamma _1}{\gamma _2}}}{{{P_S}{\gamma _1} + ({P_R}{\gamma _2} + 1)({P_R}{\gamma _R} + 1)}} = \frac{{{X_1}{P_R}{\gamma _2}}}{{{X_1} + {P_R}{\gamma _2} + 1}},
\end{equation}
where ${\gamma _1} = \frac{{|{h_1}{|^2}}}{{{\sigma ^2}}}$,${\gamma _2} = \frac{{|{h_2}{|^2}}}{{{\sigma ^2}}}$,${\gamma _R} = \frac{{|{h_{RI}}{|^2}}}{{{\sigma ^2}}}$ denote the respective channel SNRs and ${X_1} = \frac{{{P_S}{\gamma _1}}}{{{P_R}{\gamma _R} + 1}}$.
%The expression $\gamma _F$ is similar to the SINR expression of the two hop communication system in \cite{riihonen2009feasibility}.

In the HD mode, the relay R receives the signal from the source at the first half of a time slot, and it is given by
\begin{equation}
{y_r} = {h_1}\sqrt {{P_S}} x + {n_1},
\end{equation}

At second half of a time slot, relay R transmits the received signal to the destination D with AF protocol. The received signal at destination D is given by
\vspace{-2mm}
\begin{eqnarray}
{y_d} &=& {h_2}\sqrt {{P_R}}{x_r} + {n_2},\\
{x_r} &=& {\beta _h}{y_r},
\end{eqnarray}
where ${\beta _h}$ is the amplification factor. Under transmit power constraint at relay R, ${\beta _h}$ can be expressed as
\begin{equation}
{\beta _h}^2 = \frac{1}{{|{h_1}{|^2}{P_s} + {\sigma ^2}}}.
\end{equation}

At destination D, the end-to-end SINR is thus given by
\begin{equation}\label{hdsinr}
{\gamma _H}{\rm{ }} = \frac{{{P_S}{P_R}{\gamma _1}{\gamma _2}}}{{{P_S}{\gamma _1} + {P_R}{\gamma _2} + 1}}.
\end{equation}

The instantaneous SNRs ${\gamma _1}$ , ${\gamma _2}$ are modeled as the exponential random variable with respective means ${\lambda _1}$ and ${\lambda _2}$.
In the X-duplex relay system, the self interference at relay is mitigated with effective self interference cancellation techniques~\cite{duarte2010Fullduplex,duarte2012experiment,day2012full,Everett2011empower,MJain2011,EAryafar2012,riihonen2010residual,riihonen2011mitigation,senaratne2011beamforming,everett2014passive,sabharwal2013band}.
The residual self interference at relay is assumed to follow the Rayleigh distribution\cite{duarte2012experiment}.
At the relay R, the SNR of residual self interference ${\gamma _R}$ follows the exponential distribution with mean value ${\lambda _R}$.
The residual self interference level is denoted as $\eta = \frac{{{\lambda _R}{P_R}}}{{{\lambda _1}{P_S}}}$.
As the source signal might behave as interference to the self interference cancellation in active self interference cancellation schemes, the value of $\eta$ might vary with ${P_S}{\gamma _1}$.
If only passive cancellation is applied, $\eta$ might be independent to ${P_S}{\gamma _1}$.
In this paper, $\eta$ is merely used to denote the ratio of the average power of residual self interference ${\lambda _R}{P_R}$ and received signal at relay ${\lambda _1}{P_S}$,  and is not assumed to be constant.
%in active cancellation process, we use $\hat h_{RI}\sqrt {{P_R}} \hat{x_r}$ to denote the affected RSI.
%If only passive suppression is adopted, $\hat {{h_{RI}}}\hat {{x_r}} = {h_{RI}}{x_r}$.
%While in other cases,  $E[| {\hat {{h_{RI}}}\hat {{x_r}}}{|^2}]  > E[| {{h_{RI}}{x_r}}{|^2}] $.
%In this paper, RSI level $\eta = \frac{{{\lambda _R}{P_R}}}{{{\lambda _1}{P_S}}}$ is considered as a lower bound of $\hat {\eta} = \frac{{{\hat \lambda _R}{P_R}}}{{{\lambda _1}{P_S}}}$, where $\hat \lambda_R$ is the mean value of $\hat {\gamma _R} = \frac{{|\hat {h_{RI}}{|^2}}}{{{\sigma ^2}}}$.

%$\eta$ is given by
%\begin{equation}
%{\lambda _R} = \eta \lambda.
%\end{equation}

\subsection{X-duplex Relay}
The HD mode outperforms the FD mode in the severe self interference region.
To optimize the system performance, we consider a X-duplex relay which can be reduced to either FD or HD with different RF chain configurations based on the instantaneous SINR.
%In this case, we assume that the relay and source are under the same power constraint, ${P_S} = {P_R} = 1$.
The CSI of the self interference ${h_{RI}}$ can be measured by sufficient training \cite{Bpday2012}\cite{Tmkim2013}. The CSI of ${h_1}$ and ${h_2}$ can be obtained through pilot-based channel estimation.
We also assume that reliable feedback channels are deployed, therefore the CSIs can be transmitted to the decision node.

The system's average sum rate under FD and HD modes can be expressed as
\begin{equation}\label{rate}
{R_{FD}} = {\log _2}({\gamma _F} + 1),
{R_{HD}} = {\log _2}(\sqrt {{\gamma _H + 1}} ),
\end{equation}
where ${\gamma _F}$ , ${\gamma _H}$ denotes the SINR of the FD and HD modes, respectively.

To maximize the instantaneous sum rate, the instantaneous SINR of X-duplex relay can be given by
\begin{equation}
{\gamma _{\max }} = \max \{ {\gamma _F},\sqrt {{\gamma _H + 1}} - 1 \}.
\end{equation}

\subsection{Adaptive Power Allocation}
In order to further optimise the system performance, we introduce the adaptive power allocation (PA) in the X-duplex relay to maximize the relay system's end-to-end SINR subject to the total transmit power constraint, ${P_S} + {P_R} = P$.
The optimal PA scheme for FD mode and HD mode based on the instantaneous CSIs is given by\cite{riihonen2011hybrid}
\begin{align}\label{pafdhd}
&{P_{S,FD\_PA}} =\hspace{-1mm}  \frac{{\sqrt {P{\gamma _1} + 1} }}{{\sqrt {P{\gamma _1} + 1}  +\hspace{-1mm} \sqrt {\left( {P{\gamma _2} + 1} \right)\left( {P{\gamma _R} + 1} \right)} }}P, {P_{S,HD\_PA}} =\hspace{-1mm} \frac{{\sqrt {P{\gamma _1} + 1} }}{{\sqrt {P{\gamma _1} + 1} +\hspace{-1mm} \sqrt {P{\gamma _2} + 1} }}P,\nonumber\\
&{P_{R,FD\_PA}} =\hspace{-1mm} \frac{{\sqrt {( {P{\gamma _2} + 1} )( {P{\gamma _R} + 1} )} }}{{\sqrt {P{\gamma _1} + 1}  +\hspace{-1mm} \sqrt {\left( {P{\gamma _2} + 1} \right)\left( {P{\gamma _R} + 1} \right)} }}P,
{P_{R,HD\_PA}} =\hspace{-1mm} \frac{{\sqrt {P{\gamma _2} + 1} }}{{\sqrt {P{\gamma _1} + 1} +\hspace{-1mm} \sqrt {P{\gamma _2} + 1} }}P.
\vspace{-2mm}
\end{align}

Based on (\ref{pafdhd}), the respective end-to-end SINR of FD and HD modes with PA are derived as
\begin{eqnarray}
&&{\gamma _{fd\_pa}} = \frac{{{P^2}{\gamma _1}{\gamma _2}}}{{P({\gamma _1} + {\gamma _2} + {\gamma _R}) + 2 + 2\sqrt {(P{\gamma _1} + 1)(P{\gamma _2} + 1)(P{\gamma _R} + 1)} }},\nonumber\\
  &&{\gamma _{hd\_pa}} = \frac{{{P^2}{\gamma _1}{\gamma _2}}}{{P({\gamma _1} + {\gamma _2})+ 2 + 2\sqrt {(P{\gamma _1} + 1)(P{\gamma _2} + 1)} }}.
\end{eqnarray}

Therefore, the instantaneous SINR of X-duplex relay with PA can be given by
\begin{equation}\label{xdpa1}
  {\gamma _{xd\_pa}} = \max \{ {\gamma _{fd\_pa}},\sqrt {{\gamma _{hd\_pa}} + 1} -1 \}.
\end{equation}
\section{Performance Analysis}
In this section, we present the CDF of the X-duplex relay and analyze the performance of the X-duplex system, including the outage probability, SER and the average sum rate.
The derived expressions of performance of X-duplex with one shared antenna are essentially equivalent to the conventional system with two separated antennas [21].s
 \begin{lemma}\label{lemmafdcdf}
 The asymptotic complementary CDF of $\gamma_F$ is given by%, defined as $P(\gamma_F>x)=1-P(\gamma_F<x)$, s
 \begin{equation}\label{fdcdf}
\Pr ({\gamma _F} > x) \approx \frac{{{\beta _1}}}{{1 + \eta x}}{K_1}({\beta _1}){e^{ - Cx}} - \frac{{2\eta ({x^2} + x)}}{{\lambda _2}{P_R}{{{(1 + \eta x)}^2}}}{K_0}({\beta _1}){e^{ - Cx}},
 \end{equation}
 where $C = \left( {\frac{1}{{{\lambda _1}{P_S}}} + \frac{1}{{{\lambda _2}{P_R}}}} \right)$, ${\beta _1} = 2\sqrt {\frac{{x + {x^2}}}{{{\lambda _1}{\lambda _2}{P_S}{P_R}}}}$, ${K_1}( \cdot )$ , ${K_0}( \cdot )$ are the first and zero order Bessel function of the second kind \cite{abramowitz1964handbook}.
 \end{lemma}
 \begin{IEEEproof}
 The derivation is presented in Appendix A.
 \end{IEEEproof}

\begin{lemma}\label{lemmahdcdf}
The complementary CDF of $\sqrt {{\gamma _H} + 1 } - 1 $ is given by%probability of $\{{\gamma _H} > {x^2}\} $
 \begin{equation}\label{hdcdf}
\Pr \left( {{\gamma _H} > {x^2} + 2x} \right) = \frac{1}{{{\lambda _2}}}\int\limits_{({x^2} + 2x)/{P_R}}^\infty  {{e^{ - \frac{1}{{{P_S}{\lambda _1}}}({x^2} + 2x + \frac{{{{({x^2} + 2x)}^2} + {x^2} + 2x}}{{{P_R}{\gamma _2} - {x^2} - 2x}}) - \frac{1}{{{\lambda _2}}}{\gamma _2}}}d{\gamma _2}} {\rm{ = }}{\beta _2}{K_1}({\beta _2}){e^{ - C({x^2} + 2x)}},
 \end{equation}
 where ${\beta _2} = 2\sqrt {\frac{{{{({x^2} + 2x)}^2} + {x^2} + 2x}}{{{\lambda _1}{\lambda _2}{P_S}{P_R}}}}  $ .
\end{lemma}
\begin{IEEEproof}
 The HD mode's end-to-end SINR is given in (12),  with the help of \cite[eq.(3.324.1)]{zwillinger2014table}, (\ref{hdcdf}) can be obtained.
\end{IEEEproof}

\begin{lemma}\label{lemmahybridcdf}
The asymptotic probability of $\{ {\gamma _F} > x,{\gamma _H} > {x^2 + 2x}\} $ can be obtained as
 \begin{equation}
\Pr ({\gamma _F} > x,{\gamma _H} > {x^2 + 2x}){\rm{ = }}{I_1} + {I_2},
%)\mathrm H(x - 1) + {I_3}\mathrm H(1 - x),
 \end{equation}
where ${I_1}$, ${I_2}$ are expressed as
\setlength{\arraycolsep}{0.0em}
\begin{eqnarray}\label{I1}
{I_1} &&= \frac{{{\beta _3}}}{{1 + \eta x}}{K_1}({\beta _3}){e^{ - {\beta _4}}} - \frac{{2\eta ({x^2} + x)}}{{{\lambda _2}{P_R}{{(1 + \eta x)}^2}}}{K_0}({\beta _3}){e^{ - {\beta _4}}},\\\label{I2}
{I_2} &&= {\beta _2}{K_1}({\beta _2}){e^{ - C(x^2 + 2x)}} - {\beta _3}{K_1}({\beta _3}){e^{ - {\beta _4}}},
%{I_3} &&= \frac{{{\beta _1}}}{{1 + \eta x}}{K_1}({\beta _1}){e^{ - {\beta _0}}} - \frac{{2\eta ({x^2} + x)}}{{{\lambda _2}{P_R}{{(1 + \eta x)}^2}}}{K_0}({\beta _1}){e^{ - {\beta _0}}},
\end{eqnarray}
where ${\beta _3} = 2\sqrt {\frac{{{{({x^2} + 2x)}^2} + {x^2} + 2x + \frac{1}{\eta }(x + 1)({x^2} + 2x)}}{{{\lambda _1}{\lambda _2}{P_S}{P_R}}}}  $ , ${\beta _4} = C({x^2} + 2x) + \frac{{x + 1}}{{\eta {\lambda _1}{P_S}}}$.
\end{lemma}
\begin{IEEEproof}
The derivation is presented in Appendix B.
\end{IEEEproof}

\subsection{Distribution of the Received Signal}
\begin{proposition}
The asymptotic CDF of X-duplex relay system's SINR ${\gamma _{\max }}$ can be derived as
\vspace{-7mm}
\setlength{\arraycolsep}{0.0em}
\begin{eqnarray}\label{cdf}
\Pr ({\gamma _{\max }} < x) && = 1 - \frac{1}{{1 + \eta x}}\left[ {{\beta _1}{K_1}({\beta _1}){e^{ - Cx}} + \eta x{\beta _3}{K_1}({\beta _3}){e^{ - {\beta _4}}}} \right]\nonumber\\
&& + \frac{{2\eta ({x^2} + x)}}{{{\lambda _2}{P_R}{{(1 + \eta x)}^2}}}\left[ {{K_0}({\beta _1}){e^{ - Cx}} - {K_0}({\beta _3}){e^{ - {\beta _4}}}} \right],
\vspace{-2mm}
\end{eqnarray}
where $\eta  = \frac{{{\lambda _R}{P_R}}}{{{\lambda _1}{P_S}}}$, $C=  {\frac{1}{{{\lambda _1}{P_S}}} + \frac{1}{{{\lambda _2}{P_R}}}} $, ${\beta _1} = 2\sqrt {\frac{{x + {x^2}}}{{{\lambda _1}{\lambda _2}{P_S}{P_R}}}}$, ${\beta _2} = 2\sqrt {\frac{{{{({x^2} + 2x)}^2} + {x^2} + 2x}}{{{\lambda _1}{\lambda _2}{P_S}{P_R}}}}$ , ${\beta _3} = 2\sqrt {\frac{{{{({x^2} + 2x)}^2} + {x^2} + 2x + \frac{1}{\eta }(x + 1)({x^2} + 2x)}}{{{\lambda _1}{\lambda _2}{P_S}{P_R}}}}  $ , ${\beta _4} = C({x^2} + 2x) + \frac{{x + 1}}{{\eta {\lambda _1}{P_S}}}$, ${K_1}( \cdot )$, ${K_0}( \cdot )$  are the first and zero order Bessel function of the second kind.
\end{proposition}

\begin{IEEEproof}
According to the permutation theorem, the CDF expression can be obtained as
\vspace{-5mm}
\setlength{\arraycolsep}{0.0em}
\begin{eqnarray}
\Pr ({\gamma _{\max }} < x) &&= \Pr ({\gamma _F} < x,\sqrt {{\gamma _H + 1}} - 1  < x)\nonumber\\
&&= 1 - \Pr ({\gamma _F} > x) - \Pr ({\gamma _H} > {x^2 + 2x})+ \Pr ({\gamma _F} > x,{\gamma _H} > {x^2 + 2x}).
\end{eqnarray}

\hspace{3mm}With the help of Lemma \ref{lemmafdcdf}, Lemma \ref{lemmahdcdf}, Lemma \ref{lemmahybridcdf}, (\ref{cdf}) is derived.
\end{IEEEproof}
\vspace{-4mm}
\subsection{Outage Probability}
The outage probability can be given as
\vspace{-3mm}
\begin{equation}\label{outage}
{P_*} = \Pr ({\log _2}(1 + SINR) < {R_0}) = {F_{{\gamma}}}(T),
\end{equation}
where the threshold of the outage probability is set to ensure the transmit rate over $R_0$ bps/Hz, $T = {2^{{R_0}}} - 1$ and ${F_\gamma }( \cdot )$ is CDF of the end-to-end SINR $\gamma $.

The X-duplex relay configures the antenna to provide the maximum sum rate of the relay network.
With the CDF expression in (\ref{cdf}) and (\ref{outage}), the outage probability of the X-duplex relay system can be derived.

From Lemma \ref{lemmafdcdf} and (\ref{outage}), the outage probability of the FD mode can be obtained. According to \cite[eq.(10.30)]{Frank}, in the high SNR condition, when $z$ comes close to zero, the ${K_1}(z)$ function converges to $\frac{1}{z}$, and the value of ${K_0}(z)$ is comparatively small. Therefore, in the high SNR scenarios, the FD mode's outage probability is approximately given by
\begin{equation}\label{fdoutage}
{P_{out\_FD}}(x) \approx 1 - \frac{1}{{1 + \eta x}}{e^{ - Cx}},
\end{equation}
when the SNR goes infinite, the outage probability of FD mode will approach
\begin{equation}\label{FDoutfloor}
P_{out\_FD}^\infty (x) = \frac{{\eta x}}{{1 + \eta x}}.
\end{equation}

Therefore, the outage probability of FD mode is limited by the error floor which is caused by self interference at high SNR.

By substituting (\ref{cdf}) into (\ref{outage}), the outage probability of X-duplex relay system can be obtained. In the high SNR, the outage probability can be derived using the similar approximation in (\ref{fdoutage}),
\begin{equation}\label{outage1}
P_{out\_XD}^\infty (x) \approx 1 - \frac{1}{{1 + \eta x}}{e^{ - Cx}} - \frac{{\eta x}}{{1 + \eta x}}{e^{ - {\beta _4} }},
\end{equation}
when the SNR goes infinite, the outage probability of X-duplex relay system approaches to zero, indicating that there is no performance floor for X-duplex relay system in the high SNR region.

For the X-duplex relay system, the finite diversity order of SNR is provided by \cite{Narasimhan2005}
\begin{equation}
d(\lambda ) =  - \frac{{\partial \ln {P_{out}}(\lambda )}}{{\partial \ln \lambda }} =  - \frac{\lambda }{{{P_{out}}(\lambda )}}\frac{{\partial {P_{out}}(\lambda )}}{{\partial \lambda }},
\end{equation}
where ${{P_{out}}(\lambda )}$ is the system's outage probability at average SNR ${\lambda}$. We use this equation to calculate the diversity order of X-duplex relay system.

We assume the transmit power of the source and relay is the same under fixed power allocation condition, $P_S = P_R = P_t$.
The diversity order ${d_{XD}}$ of the X-duplex relay system is given as
\begin{eqnarray}\label{dxdappro}
{d_{XD}} = \frac{1}{{{P_t}}}\frac{{\frac{x}{{1 + \eta x}}(\frac{1}{{{\lambda _1}}} + \frac{1}{{{\lambda _2}}}){e^{ - Cx}} + \frac{{\eta x}}{{1 + \eta x}}\left[ {(\frac{1}{{{\lambda _1}}} + \frac{1}{{{\lambda _2}}})({x^2} + 2x) + \frac{{x + 1}}{{\eta {\lambda _1}}}} \right]{e^{ - {\beta _4}}}}}{{1 - \frac{1}{{1 + \eta x}}{e^{ - Cx}} - \frac{{\eta x}}{{1 + \eta x}}{e^{ - {\beta _4}}}}}.
\end{eqnarray}

Furthermore, the diversity order ${d_{XD}}$ can be estimated by using the Taylor's formula in \cite[eq.(1.211)]{zwillinger2014table} in the high transmit power scenario
 \begin{equation}\label{dxd}
{d_{XD}} \approx \frac{{{C_1}x + \eta x{C_1}({x^2} + 2x) + \eta x\frac{{x + 1}}{{\eta {\lambda _1}}} - \frac{1}{{{P_t}}}\left[ {{{({C_1}x)}^2} + \eta x{{\left( {{C_1}({x^2} + 2x) + \frac{{x + 1}}{{\eta {\lambda _1}}}} \right)}^2}} \right]}}{{{C_1}x + \eta x{C_1}({x^2} + 2x) + \eta x\frac{{x + 1}}{{\eta {\lambda _1}}}}},
 \end{equation}
 where ${C_1} = \frac{1}{{{\lambda _1}}} + \frac{1}{{{\lambda _2}}}$. When the transmit power goes infinite, the diversity order of X-duplex relay system approaches to one, indicating that there is no error floor in the system.

For the HD mode, from equation (\ref{rate}), the HD mode's equivalent SINR in one time slot is given as $\sqrt {{\gamma _H} + 1} - 1$.
Therefore, the outage probability of HD mode can be obtained with (\ref{hdcdf})
 \begin{equation}\label{outhd}
{P_{out\_HD}}(x) = 1- {\beta _2}{K_1}({\beta _2}){e^{ - C({x^2} + 2x)}} \approx 1 - {e^{ - C(x^2 + 2x)}}.
 \end{equation}

The finite-SNR diversity orders of FD and HD mode can be written as
 \begin{eqnarray}\label{dhd}
 {d_{FD}} &=& \frac{1}{{{P_t}}}\frac{{(\frac{1}{{{\lambda _1}}} + \frac{1}{{{\lambda _2}}})\frac{x}{{1 + \eta x}}{e^{ - Cx}}}}{{1 - \frac{1}{{1 + \eta x}}{e^{ - Cx}}}} \approx \frac{{1 - \frac{x}{{{P_t}}}(\frac{1}{{{\lambda _1}}} + \frac{1}{{{\lambda _2}}})}}{{1 + {P_t}\eta \frac{{{\lambda _1}{\lambda _2}}}{{{\lambda _1} + {\lambda _2}}}}},\nonumber\\
{d_{HD}} &=& \frac{1}{{{P_t}}}\frac{{(\frac{1}{{{\lambda _1}}} + \frac{1}{{{\lambda _2}}})({x^2} + 2x) \cdot {e^{ - C({x^2} + 2x)}}}}{{1 - {e^{ - C({x^2} + 2x)}}}} \approx 1 - \frac{1}{{{P_t}}}(\frac{1}{{{\lambda _1}}} + \frac{1}{{{\lambda _2}}})({x^2} + 2x),
 \end{eqnarray}
At medium SNR and low residual self interference, the diversity order of FD can be approximated as $ {d_{FD}} \approx {1 - \frac{x}{{{P_t}}}(\frac{1}{{{\lambda _1}}} + \frac{1}{{{\lambda _2}}})}$.
With optimal self interference cancellation, $d_{FD}$ approaches one in high SNR region.
When the SNR goes infinite, the diversity order of the FD and HD mode approaches to zero and one respectively,  indicating that the outage probability curve of X-duplex relay system is parallel with HD mode when SNR reaches this region.

The outage probability intersection of FD and HD mode can be calculated as%with (\ref{fdoutage}) and (\ref{outhd}).
\begin{equation}\label{outageintersection}
{P_t}^ * = ( {\frac{1}{{{\lambda _1}}} + \frac{1}{{{\lambda _2}}}} )\frac{{{x^2} + x}}{{\ln (1 + \eta x)}},
\end{equation}
%At the intersection point, the outage probability benefit of X-duplex scheme compared with FD mode is given as
%\begin{equation}
%  \kappa _{out}^* = \frac{{P_{out\_XD}^{\rm{*}}}}{{P_{out\_FD}^{\rm{*}}}} = 1 - \frac{{\eta x{e^{ - \frac{1}{{P_t^*}}\left[ {(\frac{1}{{{\lambda _1}}} + \frac{1}{{{\lambda _2}}})({x^2} + 2x) + \frac{{x + 1}}{{\eta {\lambda _1}}}} \right]}}}}{{1 + \eta x - {e^{ - \frac{x}{{P_t^*}}(\frac{1}{{{\lambda _1}}} + \frac{1}{{{\lambda _2}}})}}}},
%\end{equation}
when $P_t  < P_t ^ *$, the outage probability of FD is lower than HD. The intersection point is affected by self interference level $\eta$. When $\eta$ reaches zero, the intersection point goes infinite, indicating that FD outperforms HD in all SNR circumstances with ideal self interference cancellation.
\subsection{Average SER Analysis}
For linear modulation formats, the average SER can be computed as \cite{wc}
\begin{equation}\label{ser}
\overline {SER}  = {a_1}{\mathbb{E}}[Q(\sqrt {2{a_2}\gamma } )] = \frac{{{a_1}\sqrt {{a_2}} }}{{2\sqrt \pi  }}\int\limits_0^\infty  {\frac{{{e^{ - {a_2}\gamma }}}}{{\sqrt \gamma  }}{F_\gamma }(\gamma )d\gamma },
\end{equation}
where ${F_\gamma }( \cdot )$ is the CDF of $\gamma $, and $Q( \cdot )$ is the Gaussian Q-Function \cite{zwillinger2014table}. The parameters $({a_1},{a_2})$ denote the modulation formats, e.g., ${a_1} = 1,{a_2} = 1$ for the binary phase-shift keying (BPSK) modulation \cite[eq.(6.6)]{wc}.
\begin{proposition}\label{serpro}
The asymptotic average SER of the X-duplex relay system can be derived as
\begin{align}\label{serfinal}
 &\overline {SER} \approx \frac{{{a_{\rm{1}}}\sqrt {{a_2}} }}{{2\sqrt \pi  }}\left\{ {{a_2}^{ - \frac{1}{2}}\Gamma (\frac{1}{2}) - \frac{1}{{\sqrt \eta  }}{e^{\frac{1}{\eta }({a_2} + C)}}\Gamma (\frac{1}{2})\Gamma (\frac{1}{2},\frac{1}{\eta }({a_2} + C))} \right.\\
 & - \eta {e^{ - \frac{1}{{{\lambda _1}{P_S}\eta }}}}{\left( {2C} \right)^{ - \frac{3}{4}}}\Gamma (\frac{3}{2})\exp (\frac{{{\mu _1}^2}}{{8C}}){D_{ - \frac{3}{2}}}(\frac{{{\mu _1}}}{{\sqrt {2C} }})\left. { - \frac{1}{2}{\eta ^3}{e^{ - \frac{1}{{{\lambda _1}{P_S}\eta }}}}{{\left( {2C} \right)}^{ - \frac{7}{4}}}\Gamma (\frac{7}{2})\exp (\frac{{{\mu _2}^2}}{{8C}}){D_{ - \frac{7}{2}}}(\frac{{{\mu _2}}}{{\sqrt {2C} }})} \right\},\nonumber
\end{align}
where ${\mu _1} = {a_2} + 2C + \frac{1}{{{\lambda _1}{P_S}\eta }} + \eta $, ${\mu _2} = {a_2} + 2C + \frac{1}{{{\lambda _1}{P_S}\eta }} + \frac{5}{3}\eta $, $\Gamma ( \cdot )$ is the Gamma Function, $\Gamma (a,x)$ is the incomplete Gamma Function, ${D_{p}}( \cdot )$ is the Parabolic Cylinder Function \cite{zwillinger2014table}.
%$C = \frac{1}{{{\lambda _1}{P_S}}} + \frac{1}{{{\lambda _2}{P_R}}}$, $\alpha  = \frac{\eta }{{\eta  + 1}}$ , ${\mu _1} = C + {a_2} + \alpha ,$ , ${\mu _2} = C + {a_2} + \frac{5}{3}\alpha $ , ${\mu _3} = 2C + \frac{1}{{{\lambda _1}{P_S}\eta }} + {a_2} + \alpha $ , ${\mu _4} = 2C + \frac{1}{{{\lambda _1}{P_S}\eta }} + {a_2} + \frac{5}{3}\alpha $, $\Gamma ( \cdot )$ is the Gamma Function, $\Gamma (a,x)$ , $\gamma (a,x)$ are the incomplete Gamma Functions and ${W_{\lambda ,\mu }}(z)$ is the Whittaker function \cite{zwillinger2014table}.
\end{proposition}

\begin{IEEEproof}
The derivation is presented in Appendix C.
\end{IEEEproof}

According to (\ref{cdf}), when SNR goes infinite, the CDF of $\gamma _{\max }$ becomes $\Pr ({\gamma _{\max }} < x) = 0$, the SER of X-duplex relay system comes to zero.

For the FD mode and HD mode, the average SER can be given as
\begin{eqnarray}
  {\overline {SER} _{FD}} &&\approx \frac{{{a_{\rm{1}}}\sqrt {{a_2}} }}{{2\sqrt \pi  }}\left\{ {{a_2}^{ - \frac{1}{2}}\Gamma (\frac{1}{2}) - \frac{1}{{\sqrt \eta  }}{e^{\frac{1}{\eta }({a_2} + C)}}\Gamma (\frac{1}{2})\Gamma (\frac{1}{2},\frac{1}{\eta }({a_2} + C))} \right\}, \nonumber\\
  {\overline {SER} _{HD}} &&\approx \frac{{{a_{\rm{1}}}\sqrt {{a_2}} }}{{2\sqrt \pi  }}\left\{ {{a_2}^{ - \frac{1}{2}}\Gamma (\frac{1}{2}) - {{(2C)}^{ - \frac{1}{2}}}\Gamma (\frac{1}{2})exp(\frac{{{{({a_2} + 2C)}^2}}}{{8C}}){D_{ - \frac{1}{2}}}(\frac{{{a_2} + 2C}}{{\sqrt {2C} }})} \right\}.
\end{eqnarray}

For the FD mode, when SNR goes infinite, the CDF of FD mode approaches $\Pr ({\gamma _{FD}} < x) = 1 - \frac{1}{{1 + \eta x}}$. With (\ref{ser}) and \cite[eq.(3.383.10)]{zwillinger2014table}, the SER of FD mode can be obtained.
\begin{equation}\label{serlowerbound}
  {\overline {SER} _{FD\_SNR \to \infty }} = \frac{{{a_1}\sqrt {{a_2}} }}{{2\sqrt \pi  }}\int\limits_0^\infty  {\frac{{{e^{ - {a_2}x}}}}{{\sqrt x }}(1 - \frac{1}{{1 + \eta x}})dx } = \frac{{{a_1}\sqrt {{a_2}} }}{{2\sqrt \pi  }}{\left( {\frac{1}{\eta }} \right)^{1/2}}{e^{\frac{1}{\eta }{a_2}}}\Gamma (\frac{3}{2})\Gamma ( - \frac{1}{2},\frac{1}{\eta }{a_2}).\nonumber \\
\end{equation}

From (\ref{serlowerbound}), it can be seen that the SER of FD mode is restricted by the lower bound, determined by self interference level $\eta$, $a_1$, $a_2$. Compared with FD mode, the X-duplex relay system removes the error floor and achieves lower SER in high SNR region.
\subsection{Average Sum Rate}
By using the CDF of ${\gamma _{\max }}$,  the average sum rate of X-duplex system is derived in this section.
\begin{equation}\label{sumrate-origin}
\bar R = {\mathbb{E}}[{\log _2}(1 + \gamma )] = \frac{1}{{\ln 2}}\int\limits_0^\infty  {\frac{{1 - {F_\gamma }(x)}}{{1 + x}}} dx,
\end{equation}
where ${F_\gamma }( \cdot )$ is the CDF of $\gamma $.

In order to simplify the final average sum rate expression, ${w_{i1}}(a,b)$ and ${w_{i2}}$ are introduced to denote the approximate value of integral $\int_b^\infty  {{e^{ - C{x^2}}}/(x + a)dx} $ and $\int_{1 + \frac{1}{{4\eta }}}^{1 + \frac{5}{{4\eta }}} {{e^{ - C{x^2}}}/(x + \frac{3}{{4\eta }})dx}$, given in Lemma \ref{sumrateintegral} and \ref{lemmaw3p}.
\begin{lemma}\label{sumrateintegral}
when $\left| b \right| > \left| a \right| > 0$, the exact value of integral ${w_{i1}}(a,b) = \int_b^\infty  {{e^{ - C{x^2}}}/(x + a)dx} $  is given by
\begin{small}\begin{equation}\label{wi1}
{w_{i1}}(a,b) = \frac{1}{{2a}}\sqrt {\frac{\pi }{C}} \left[ {1 - \Phi (C{b^2})} \right] + \frac{{{e^{ - C{a^2}}}}}{2}{E_1}(C({b^2} - {a^2}))
  - \frac{{{e^{ - \frac{{C{b^2}}}{2}}}}}{{2a}}\sum\limits_{k = 1}^\infty  {{a^{2k - 2}}{C^{k - \frac{3}{2}}}{{(C{b^2})}^{\frac{1}{4} - \frac{k}{2}}}{W_{\frac{1}{4} - \frac{k}{2},\frac{3}{4} - \frac{k}{2}}}(C{b^2})},
\end{equation}\end{small}
%\begin{eqnarray}\label{wi1}
%{w_{i1}}(a,b) =&& \frac{1}{{2a}}\sqrt {\frac{\pi }{C}} \left[ {1 - \Phi (C{b^2})} \right] + \frac{{{e^{ - C{a^2}}}}}{2}{E_1}(C({b^2} - {a^2}))\nonumber\\
% && - \frac{{{e^{ - \frac{{C{b^2}}}{2}}}}}{{2a}}\sum\limits_{k = 1}^\infty  {{a^{2k - 2}}{C^{k - \frac{3}{2}}}{{(C{b^2})}^{\frac{1}{4} - \frac{k}{2}}}{W_{\frac{1}{4} - \frac{k}{2},\frac{3}{4} - \frac{k}{2}}}(C{b^2})} ,
% \end{eqnarray}
where $\Phi ( \cdot )$ is the probability integral, and  ${W_{\lambda ,\mu }}(z)$ is the Whittaker function \cite{zwillinger2014table}, we use the first ${{N}}$ items of the third part of (\ref{wi1}) for approximation, denoted as ${w_{i1}}(a,b,N)$.
\end{lemma}
\begin{IEEEproof}
The derivation is presented in Appendix D.
\end{IEEEproof}
\begin{lemma}\label{lemmaw3p}
The approximate value of integral ${w_{i2}} = \int_{ \rho }^{1 + \frac{1}{\eta }} {{e^{ - C{x^2}}}/(x + \frac{1}{\eta } - \rho )dx} $ is given by
\begin{small}\begin{equation}\label{w3p}
{w_{i2}} \approx {e^{ - C{\rho ^2} + 2C\rho \frac{1}{\eta }}}\sum\limits_{k = 0}^{{N_2}} {\frac{{{{( - C)}^k}}}{{k!{\eta ^{2k}}}}\left[ {{E_1}({\varepsilon _1}) - {E_1}({\varepsilon _2})} \right]}
+ {e^{ - C{\rho ^2} + 2C\rho \frac{1}{\eta }}}\sum\limits_{k = 1}^{{N_2}} {\sum\limits_{l = 1}^{2k} {\frac{{{{( - C)}^k}}}{{k!{{(2C\rho )}^l}{{( - \eta )}^{2k - l}}}}(\begin{array}{*{20}{c}}
{2k}\\
l
\end{array})} } \left[ {\gamma (l,{\varepsilon _2}) - \gamma (l,{\varepsilon _1})} \right],
\end{equation}\end{small}
%\setlength{\arraycolsep}{0.0em}
%\begin{eqnarray}\label{w3p}
%  {w_{i2}} \approx&& {e^{ - C{\rho ^2} + 2C\rho \frac{1}{\eta }}}\sum\limits_{k = 0}^{{N_2}} {\frac{{{{( - C)}^k}}}{{k!{\eta ^{2k}}}}\left[ {{E_1}({\varepsilon _1}) - {E_1}({\varepsilon _2})} \right]}    \nonumber\\
% && + {e^{ - C{\rho ^2} + 2C\rho \frac{1}{\eta }}}\sum\limits_{k = 1}^{{N_2}} {\sum\limits_{l = 1}^{2k} {\frac{{{{( - C)}^k}}}{{k!{{(2C\rho )}^l}{{( - \eta )}^{2k - l}}}}(\begin{array}{*{20}{c}}
%{2k}\\
%l
%\end{array})} } \left[ {\gamma (l,{\varepsilon _2}) - \gamma (l,{\varepsilon _1})} \right],
% \end{eqnarray}
where ${\varepsilon _0} = 1 + \frac{2}{\eta } - \rho$, ${ \varepsilon _1} = 2C\rho \frac{1}{\eta }$, ${\varepsilon _2} = 2C\rho {\varepsilon _0}$, $\gamma (a,x)$ is the incomplete Gamma Function,  first ${{N_2}}$ items are used to approximate value.
\end{lemma}
\begin{IEEEproof}
The derivation is presented in Appendix E.
\end{IEEEproof}

\begin{proposition}\label{sumrate}
The average sum rate of X-duplex system can be expressed approximately as
\begin{eqnarray}\label{longrate}
\bar R \approx&& \frac{1}{{\ln 2}}\left\{ \frac{{\rm{1}}}{{1 - \eta }}\left[ {{e^C}{E_1}(C) - {e^{\frac{C}{\eta }}}{E_1}(\frac{C}{\eta })} \right] -\frac{2}{{{\lambda _2}{P_R}{C_2}}}{e^{\frac{C}{{2\eta }}}}{\Gamma ^2}(2){W_{ - \frac{3}{2},0}}({z_1}){W_{ - \frac{3}{2},0}}({z_2}) \right.\nonumber\\
 &&\left. { + \frac{\eta }{{\eta  - 1}}{e^{C{\rho ^2} - \frac{1}{{{\lambda _1}{P_S}\eta }}}}\left[ {{w_{i1}}(1 - \rho ,\rho ,{N_1}) - \frac{1}{\eta }{w_{i1}}(\frac{1}{\eta } - \rho ,1 + \frac{1}{\eta },{N_3}) - \frac{1}{\eta }{w_{i2}}} \right]} \right\}.
\end{eqnarray}
where ${C_2} = \frac{2}{{\sqrt {{\lambda _1}{\lambda _2}{P_S}{P_R}} }}$, ${z_1} = \frac{{C + \sqrt {{C^2} - {{C_2}^2}} }}{{2\eta }}$, ${z_2} = \frac{{C - \sqrt {{C^2} - {{C_2}^2}} }}{{2\eta }}$, ${W_{\lambda ,\mu }}(z)$ is the Whittaker functions \cite{zwillinger2014table}.
\end{proposition}
\begin{IEEEproof}
The derivation is presented in Appendix F.
\end{IEEEproof}

According to (\ref{cdf}) and (\ref{sumrate-origin}), when SNR goes infinite, the CDF of $\gamma _{\max }$ becomes $\Pr ({\gamma _{\max }} < x) = 1 - H(x - 1) - H(1 - x) = 0$ and the average sum rate of X-duplex relay system can be derived as,
\begin{equation}\label{RATEXD}
  {{\bar R}_{XD\_SNR \to \infty }} = \frac{1}{{\ln 2}}\int\limits_0^\infty  {\frac{1}{{1 + x}}} dx.
\end{equation}

It can be observed that the maximal achievable average sum rate of X-duplex relay system is not restricted by the self interference.

The approximate average sum rate of FD mode and HD mode can be given as
\begin{equation}
  {{\bar R}_{FD}} \approx \frac{1}{{\ln 2}}\frac{{\rm{1}}}{{1 - \eta }}\left[ {{e^C}{E_1}(C) - {e^{\frac{C}{\eta }}}{E_1}(\frac{C}{\eta })}\right],
  {{\bar R}_{HD}} \approx \frac{1}{{2\ln 2}}{e^C}{E_1}(C),
\end{equation}

%Assuming $P_S =P_R =P_t$, the intersection point of the average sum rate of FD and HD can be given as
%\begin{equation}\label{conjunc}
%  {{C}}_{rate}^* = \frac{{\ln ({E_1}({{C}}_{rate}^*)) - \ln ({E_1}(\frac{{{C}}_{rate}^*}{\eta }))}}{{(1 - \frac{1}{\eta })\left[ {\ln (1 + \eta ) - \ln 2} \right]}},
%\end{equation}
%with mathematical tools, ${{P_{trate}}^*} = \frac{1}{{{C}}_{rate}^*}(\frac{1}{{{\lambda _1}}} + \frac{1}{{{\lambda _2}}})$ can be derived.

When SNR goes infinite, the upper bound of FD mode can be derived \cite[eq.(3.195)]{zwillinger2014table}.
\begin{equation}\label{ratebound}
  {{\bar R}_{FD\_SNR \to \infty }} = \frac{1}{{\ln 2}}\int\limits_0^\infty  {\frac{1}{{1 + x}}\frac{1}{{1 + \eta x}}} dx = \frac{{\ln \eta }}{{(\eta  - 1)\ln 2}}.
\end{equation}

The upper bound of the average sum rate of FD mode is given in (\ref{ratebound}).
It means that the practical average sum rate cannot be larger than (\ref{ratebound}), which presents the achievable region of average sum rate of FD mode.

Comparing the (\ref{RATEXD}) and (\ref{ratebound}), the X-duplex relay system overcomes the restriction of self interference compared with FD mode.

\subsection{Diversity order of XD-PA}
In this subsection, a lower bound and a upper bound for the X-duplex relay's end-to-end SINR with PA are provided and the CDF of these bounds are obtained. Finally, the diversity order of XD-PA is derived.

The lower bound and upper bound for the end-to-end SINR (\ref{xdpa1}) can be written as
\begin{equation}\label{bound}
  \mathcal D({\gamma _{lower}},{\gamma _R}) \ge {\gamma _{xd\_pa}} \ge \mathcal D({\gamma _{upper}},{\gamma _R}),
\end{equation}
where ${\gamma _{upper}} = \min \{ {\gamma _1},{\gamma _{\rm{2}}}\} $, ${\gamma _{lower}} = \max {\rm{\{ }}{\gamma _{\rm{1}}},{\gamma _{\rm{2}}}{\rm{\} }}$, $\mathcal D(x,y) = \max \{ \mathcal C(x,y),{\sqrt {\mathcal C(x,0) + 1} -1} \} $, $\mathcal C(x,y) = \frac{{{x^2}{P^2}}}{{2xP + yP + 2 + 2(xp + 1)\sqrt {yp + 1} }}$. When $x \in (0, + \infty )$, the function $\mathcal C(x,y)$ is a monotonically increasing function. Therefore, the function $\mathcal D(x,y)$ is also monotonic when $x \in (0, + \infty )$.

The CDF distribution of ${\gamma _{upper}}$ , ${\gamma _{lower}}$ is given as
\begin{equation}
{F_{_{{\gamma _u}}}}(x) = 1 - {e^{ - \frac{1}{{{\lambda _1}}}x - \frac{1}{{{\lambda _2}}}x}},
{F_{_{{\gamma _l}}}}(x) = (1 - {e^{ - \frac{1}{{{\lambda _1}}}x}})(1 - {e^{ - \frac{1}{{{\lambda _2}}}x}}).
\end{equation}

With \cite[eq.(3.322)]{zwillinger2014table}, we can obtain the outage probability of the upper bound $\mathcal D({\gamma _{upper}},{\gamma _R})$
\begin{eqnarray}
  {P_{{\gamma _{upper}}}}(x) &&= \Pr \{ \mathcal C({\gamma _u},{\gamma _R}) < x,\mathcal C({\gamma _u},0) < {x^2 + 2x}\} = \int_0^{\frac{2}{T}} {{f_{{\gamma _u}}}(t)dt}  + \int_{\frac{2}{T}}^{{T_2}} {{e^{ - \frac{{{T^2}{t^2} - 2Tt}}{{P{\lambda _R}}}}}{f_{{\gamma _u}}}(t)dt}  \nonumber\\
   && = {F_{{\gamma _u}}}(\frac{2}{T}) + (\frac{1}{{{\lambda _1}}} + \frac{1}{{{\lambda _2}}})G(\frac{2}{T},{T_2},\frac{{P{\lambda _R}}}{{4{T^2}}},\frac{1}{{{\lambda _1}}} + \frac{1}{{{\lambda _2}}} -  \frac{{2T}}{{P{\lambda _R}}}),
\end{eqnarray}
where  $T = \frac{{\sqrt {x + 1}  - \sqrt x }}{{\sqrt x }}P $ , ${T_2} = \frac{{2({x^2} + 2x) + 2\sqrt {{{({x^2} + 2x)}^2} + {x^2} + 2x} }}{P}$, ${f_{{\gamma _u}}}(t)$ is the PDF of ${\gamma _{upper}}$, $G({u_1},{u_2},\beta ,\gamma ) = \sqrt {\pi \beta } {e^{\beta {\gamma ^2}}}\left[ {\Phi (\gamma \sqrt \beta   + \frac{{{u_2}}}{{2\sqrt \beta  }}) - \Phi (\gamma \sqrt \beta   + \frac{{{u_1}}}{{2\sqrt \beta  }})} \right]$.

The Taylor expansion of the upper bound ${P_{{\gamma _{upper}}}}(x)$ is
\begin{equation}\label{upperbound}
 {P_{{\gamma _{upper}}}}(x){\rm{ = }}\frac{{2({\lambda _1}{\rm{ + }}{\lambda _2})}}{{{\lambda _1}{\lambda _2}}}\frac{{{x^2} + 2x + \sqrt {{{({x^2} + 2x)}^2} + {x^2} + 2x}  - \sqrt {{x^2} + x}  - x}}{P} + o({P^{ - \frac{3}{2}}}).
\end{equation}

It can be observed that the diversity order of XD-PA is at least one.

Similarly, the outage probability of the lower bound $\mathcal D({\gamma _{upper}},{\gamma _R})$ can be calculated as
\begin{eqnarray}
{P_{{\gamma _{lower}}}}(x) =&& {F_{{\gamma _l}}}(\frac{2}{T}) + \frac{1}{{{\lambda _1}}}G(\frac{2}{T},{T_2},\frac{{P{\lambda _R}}}{{4{T^2}}},\frac{1}{{{\lambda _1}}} - \frac{{2T}}{{P{\lambda _R}}}) + \frac{1}{{{\lambda _2}}}G(\frac{2}{T},{T_2},\frac{{P{\lambda _R}}}{{4{T^2}}},\frac{1}{{{\lambda _2}}} - \frac{{2T}}{{P{\lambda _R}}})\nonumber\\
    && - (\frac{1}{{{\lambda _1}}} + \frac{1}{{{\lambda _2}}})G(\frac{2}{T},{T_2},\frac{{P{\lambda _R}}}{{4{T^2}}},\frac{1}{{{\lambda _1}}} + \frac{1}{{{\lambda _2}}} - \frac{{2T}}{{P{\lambda _R}}}),
\end{eqnarray}
using Taylor's formula, we can obtain
\begin{equation}\label{lowerbound}
  {P_{{\gamma _{lower}}}}(x) = \frac{{2{\lambda _R}}}{{{\lambda _1}{\lambda _2}}}\frac{{{T_3}}}{{{P^2}}}\left( {3\frac{{{\lambda _1}{\rm{ + }}{\lambda _2}}}{{{\lambda _1}{\lambda _2}}} - \frac{8}{{{\lambda _R}}}\frac{{\sqrt {x + 1}  - \sqrt x }}{{\sqrt x }}} \right) + o({P^{ - \frac{5}{2}}}),
\end{equation}
where ${T_3} = {(\sqrt {{x^2} + x}  + x)^2}({x^2} + 2x + \sqrt {{{({x^2} + 2x)}^2} + {x^2} + 2x}  - \sqrt {{x^2} + x}  - x)$.

It can be observed that the diversity order of XD-PA is at most two.

Similarly, the Taylor expansion of the upper bound and lower bound of the outage probability of the FD mode with PA can be provided as
\vspace{-2mm}
\begin{eqnarray}
  P_{{\gamma _{upper}}}^{FD}(x) &&= \frac{{({\lambda _1}{\rm{ + }}{\lambda _2})\sqrt {{\lambda _R}\pi } }}{{2{\lambda _1}{\lambda _2}}}\frac{{\sqrt {{x^2} + x}  + x}}{{\sqrt P }} + o({P^{ - 1}}),\nonumber\\
  P_{{\gamma _{lower}}}^{FD}(x) &&= \frac{{{\lambda _R}}}{{{\lambda _1}{\lambda _2}}}\frac{{{{(\sqrt {{x^2} + x}  + x)}^2}}}{P} + o({P^{ - 2}}),
\end{eqnarray}
The diversity order of FD with PA is between $\frac{1}{2}$ and $1$.
As the diversity order of X-duplex is one at high SNR, the diversity order of X-duplex is higher than FD with PA at high SNR.

\section{Simulation Results}
In this section, simulations are provided to validate the performance analysis of the relay system with X-duplex relay.
Without loss of generality, we set the SNRs of source-relay and relay-destination channel as one, $\lambda_1 = {\lambda _2} = 1$.
The transmit power of the source and relay is set as equal under the fixed power allocation condition, $P_S = P_R$.
The threshold of the outage probability is set as 2 bps/Hz \cite{krikidis2012full,7065323}.
It is shown in \cite{MJain2011,bharadia2014full,MChung} that the self interference can be cancelled up to 110 dB.
We assume the self interference cancellation ability is between 70dB and 110dB~\cite{zzhang}.
The path loss between source and relay is modeled as $P{L_{LOS}}\left( R \right) = 103.4 + 24.2lo{g_{10}}\left( R \right)$~\cite{tr}.
Therefore, the residual self interference level $\eta$ is set as ${\eta = 0.2, 0.05, 0.01}$.

Fig. 2 demonstrates the outage probability performance of X-duplex relay system with different self interference $\eta =$ 0.2, 0.05 and 0.01.
The outage performance of FD mode and HD mode is also illustrated for comparison.
As can be seen, the exact outage probability curves tightly matches with the analytic expression given in (\ref {outage}).
The figure reveals that X-duplex relay system's outage probability is lower than both FD and HD schemes.
At high SNR, the FD scheme has an error floor, which coincides with the analytical results in (\ref{FDoutfloor}).
When the SNR goes infinite, the X-duplex relay eliminates the error floor and remains the full diversity order, as shown in (\ref {dxd}) and (\ref {dhd}).
The effect of self interference on the X-duplex relay system is very small at high SNR.
This is because the HD mode is more likely to be selected in the X-duplex relay as the performance of FD mode is interference limited at high SNR.
The X-duplex benefits more from the HD mode, whose performance is independent of residual self interference and improves with the increase of transmit power.
Therefore, the impact of residual self interference from FD mode on X-duplex becomes smaller as SNR increases and the curves of X-duplex under different $\eta$ become close.
%thus the impact of the performance floor in FD mode on X-duplex relay system is mitigated with the increase of SNR.

\begin{figure}
\centering
% Requires \usepackage{graphicx}out
\includegraphics[width=4in]{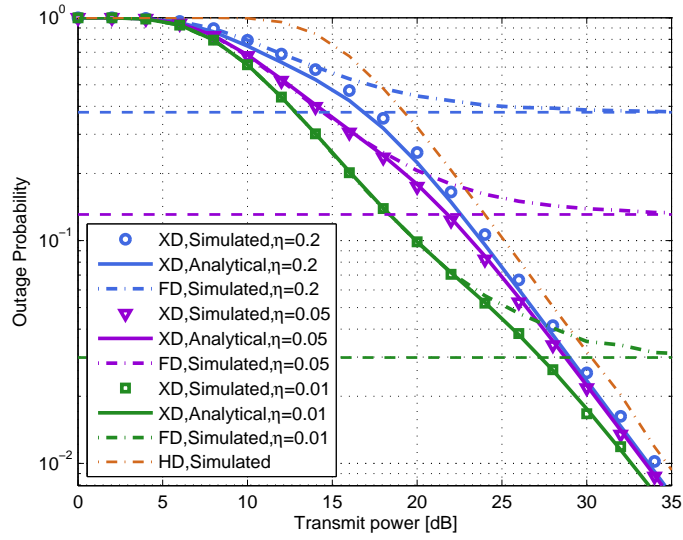}
\vspace{-2mm}
\caption{Outage probability of X-duplex relay system when $\eta {\rm{ = 0}}{\rm{.2}}$, 0.05, 0.01, the dashed lines of performance floor coincide with analytical results in (\ref{FDoutfloor}), and the intersection point of FD and HD mode coincides with analytical results in (\ref{outageintersection}).}\label{fig_outage}
\vspace{-6mm}
\end{figure}

Fig. 3 compares the finite SNR diversity order of X-duplex relay with pure FD and HD mode at $\eta = 0.2, 0.05, 0.01$.
%It can be observed that the diversity order of X-duplex is larger than that of FD mode in the whole SNR region.
The diversity order of X-duplex relay system increases with that of FD mode from low to medium SNR as FD mode is more likely to be selected in this region.
When the diversity order of FD mode decreases, the performance of X-duplex relay system is influenced.
As the performance of HD mode improves with SNR, the diversity order of X-duplex relay system increases as HD mode is more likely to be selected.
When SNR goes infinity, the diversity order curve of X-duplex relay system approaches that of HD mode because FD mode encounters the performance floor.
%Due to the existence of self interference, the finite SNR diversity order of FD mode approaches zero at high SNR.
At high SNR, the X-duplex relay eliminates the error floor and achieves the full diversity order as the HD mode, which is consistent with Section III B.

\begin{figure}[!t]
  \centering
  \includegraphics[width=4in]{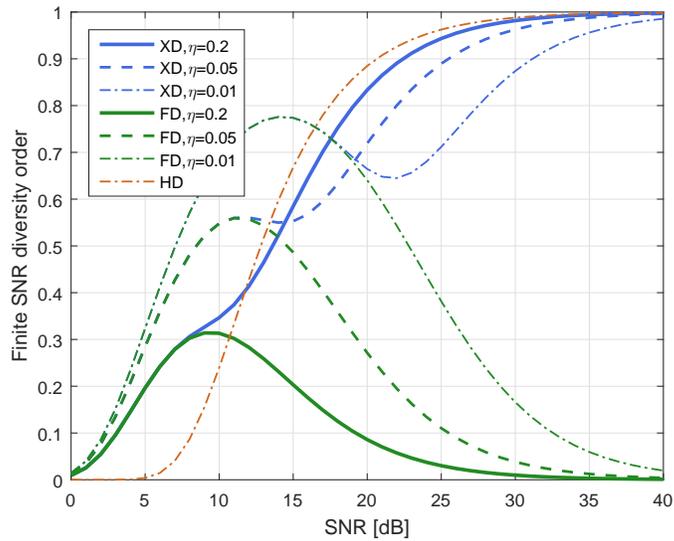}
  \vspace{-2mm}
  \caption{Finite SNR diversity order of FD mode, HD mode and X-duplex relay versus link SNR.}\label{fig_outage_cmp}
  \vspace{-6mm}
\end{figure}

\begin{figure}[!t]
  \centering
  \includegraphics[width=4in]{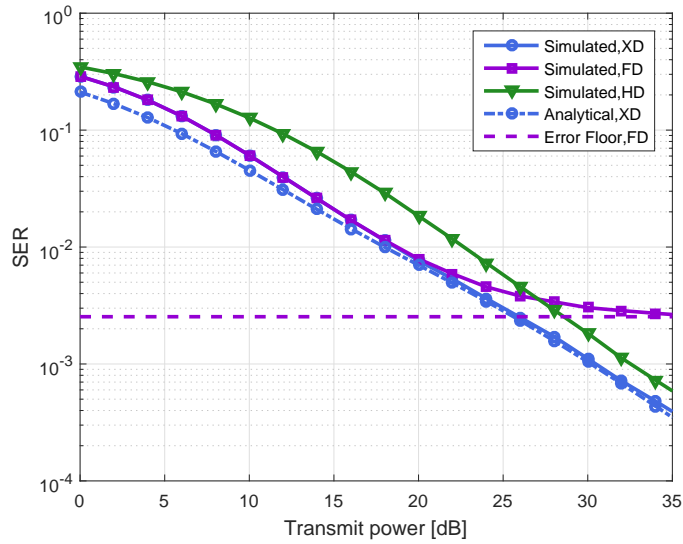}
  \vspace{-2mm}
  \caption{Average SER of X-duplex relay system when $\eta {\rm{ = 0}}{\rm{.01}}$, the dashed lines coincide with analytical results in (\ref{serlowerbound}).}\label{fig_ser}
  \vspace{-6mm}
\end{figure}

  Fig. 4 plots both the analytical and simulated results of the SER in the X-duplex relay system with $\eta {\rm{ = 0}}{\rm{.01}}$.
  The SER performance of FD and HD is depicted for comparison.
  From the figure, we can observe that X-duplex relay system achieves a better performance compared with pure FD and HD schemes.
  At high SNR , the X-duplex relay removes the performance floor.
  The curves of X-duplex and HD mode become close at high SNR as the benefit from FD mode is limited by the residual self interference.
  %Besides, X-duplex is not lower bounded by the error floor of the FD scheme in the high SNR region.

  Fig. 5 depicts the average sum rate of the X-duplex system versus SNR with $\eta  = 0.2$.
  The approximate analytical expression in (\ref{longrate}) tightly approaches the exact average sum rate.
  It can be seen from the figure that X-duplex relay system provides a higher sum rate than that of FD and HD.
  The performance improvement of X-duplex is most significant at medium SNR.

\begin{figure}[!t]
  \centering
  \includegraphics[width=4in]{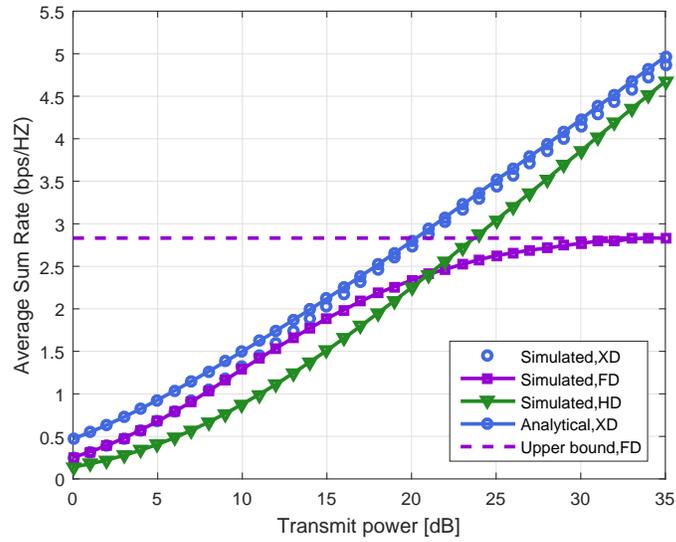}
  \vspace{-2mm}
  \caption{Average sum rate of X-duplex relay system when $\eta  = 0.2$, the dashed lines coincide with analytical results in (\ref{ratebound}).}\label{fig_sumrate}
  \vspace{-6mm}
\end{figure}

  In Fig. 6, the simulated average sum rate of the X-duplex system versus self interference with different levels of transmit power is depicted.
  In the weak self interference region, FD achieves a higher sum rate than HD.
  As self interference increases, the average sum rate of FD mode significantly decreases and performs worse than the HD mode.
  The average sum rate of X-duplex relay system is always better than FD and HD mode.
  The performance of X-duplex decreases quickly with the self interference increases, and is most obvious at high SNR.
  When the self interference is perfectly cancelled, the average sum rate of X-duplex is twice that of HD mode.

\begin{figure}[!t]
  \centering
  \includegraphics[width=4in]{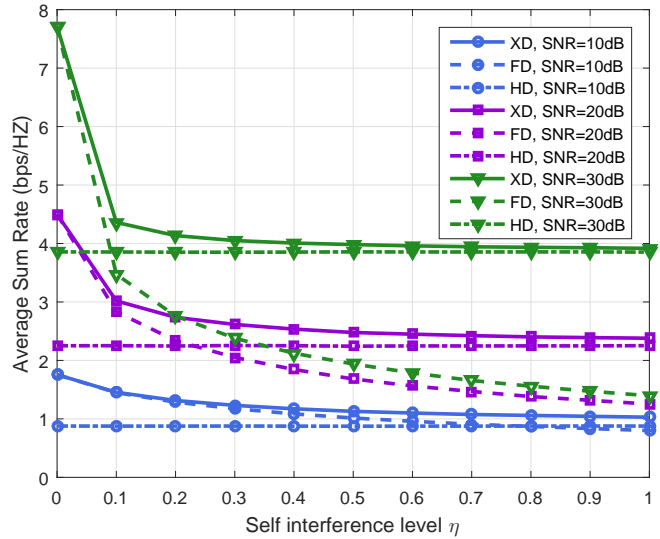}
  \vspace{-2mm}
  \caption{Average sum rate of X-duplex relay system versus self interference.}\label{fig_sumrate_SI}
  \vspace{-6mm}
\end{figure}

Fig. 7 illustrates the outage probability of XD-PA subject to the total power constraint.
The performance of the X-duplex relay system with uniform power allocation, is illustrated for comparison.
According to this figure, the outage probability performance of X-duplex relay system can be improved with adaptive power allocation compared with equal power allocation.
The diversity order of XD-PA is between one and two.
The performance of FD with power allocation is also plotted for comparison and the diversity order of FD-PA is between $\frac{1}{2}$ and one.
It can observed that the diversity order of X-duplex is higher than FD with PA at high SNR, which coincide with the analysis in Section III E.

\begin{figure}[!t]
  \centering
  \includegraphics[width=4in]{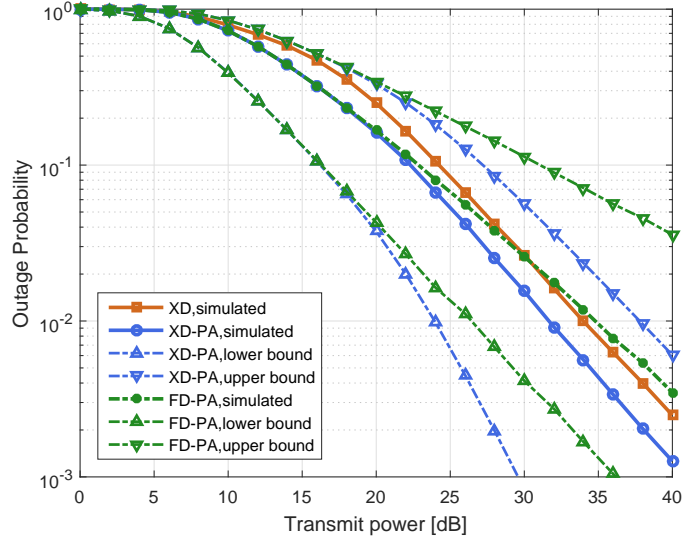}
  \vspace{-2mm}
  \caption{Outage probability of XD-PA versus the transmit power when $\eta = 0.2$.}\label{fig_pa}
  \vspace{-6mm}
\end{figure}

\subsection{Differences and Discussions}
The system model of hybrid FD/HD relaying~\cite{riihonen2011hybrid}, RAMS scheme~\cite{7065323} and X-duplex in this paper can be classified into three categories according to the deployment of antennas at the relay:
(a) Separated antenna without antenna selection [21],
(b) Separated antenna with antenna selection [20],
(c) Shared antenna in this paper.
The major differences between these three categories can be summarized as follows:

\emph{Structure and implementation: }
In (a), (b), (c), the number of antennas at the relay are two, two and one, respectively.
The connection between the antenna and RF chain is fixed in (a), however it is flexible in (b) and (c).
%In (a), there are two possible modes where the two antennas are configured as Tx/Rx and Rx/Tx.
In (a) and (b), as the channels between the source and two antennas at relay may be different in practical scenarios, we need to determine which antenna is selected as Tx antenna, and the other as Rx antenna.
In (a), the decision is made at deployment time and the configuration of each antenna is fixed.
In (b), as each antenna can be configured as Tx or Rx antenna, the deployment of antennas is simpler compared with (a).
However, the decision system could be more complex as two antennas can be adaptively configured according to instantaneous channel information and thus more operating modes need to be considered compared with (a).
In [20], there are two FD modes where two antennas are configured as Tx/Rx or Rx/Tx. %, and two cases with both antennas are configured as Tx and both are configured as Rx.
In a shared antenna relay system (c), since Tx/Rx share one single antenna, there will be no Tx/Rx selection process involved as there is only one channel between the source and relay.

\emph{Performance: }
Compared with (a), the system in (b) can provide an additional spatial diversity gain at the destination and improve the performance with efficient utilization of two antennas.
Specifically, considering one relay, the system in (b) achieves twice of the diversity order at low to medium SNRs and a lower error floor at high SNRs compared with the fixed antenna configuration (a) operating at FD mode [20].
Comparing (c) and (a), one shared antenna can operate the same way as two separated fixed antennas.
The shared antenna can exploit antenna resources more efficiently compared with fixed antennas.
Thus, (c) is more suitable to be deployed into small equipments, such as mobile phone, small sensor nodes.
The performance of X-duplex relaying system in (c) is the same as that of hybrid FD/HD switching in (a).

\emph{Complexity: }
In (a) and (c), the CSIs of three channels, including the channel from source to relay, the channel from relay to destination and the self interference channel, need to be measured and sent to the decision node for decision through feedback channels, which requires ${\log _2}(3)$ feedback overhead.
In (b), the CSIs of 5N channels, including 2N channels from the source to N relays for two antenna modes at relay, N self-interference channel at N relays, and 2N channels from N relays to the destination for two antenna modes at relay, requires the feedback overhead of ${\log _2}(5N)$.
From this perspective, the complexity of (a) and (c) is the same and smaller than that of (b) where more CSIs need to be estimated and transmitted.

%In the X-Duplex scheme, there is one source node, one shared-antenna relay and one destination node.
%The CSIs of three channels, including the channel from source to relay, the channel from relay to destination and the self interference channel, need to be measured and sent to the decision node for decision through feedback channels.
%This requires ${\log _2}(3)$ feedback overhead.
%Based on the information, the decision node selects the optimal antenna mode for transmission.
%
%In this paper, it can be observed that with one single shared antenna relay, we can achieve one diversity order, which is equal to that of the hybrid relaying scheme where the relay is equipped with two separated antennas in \cite{riihonen2011hybrid}.
%Therefore, the shared antenna can exploit antenna resources more efficiently compared with fixed antennas.
%In addition, as the channels between the source and two antennas at the relay may be different in practical scenarios, we need to determine which antenna is chosen as receiving antenna, and the other as transmitting antenna.
%In a shared antenna relay system, such selection problem can be omitted as there is only one channel between the source and relay.
\vspace{-4mm}
\section{Conclusion}
 In this paper, we investigated a X-duplex relay for the AF relay network, in which the relay is equipped with a shared antenna.
By adaptively configuring the antenna connection with two RF chains, the X-duplex relay system can achieve a better performance than both HD and FD schemes and eliminate the performance floor of FD caused by the residual self-interference.
We also designed the XD-PA subject to the total power constraint to further improve the performance.
Asymptotic expressions of the CDF, outage probability, average SER performance, and average sum rate were derived.
The analytic results were validated by computer simulations.
Both analysis and simulations demonstrated the superiority of the X-duplex relay over both FD and HD schemes.

\appendices
\section{Proof of Lemma \ref{lemmafdcdf}}
The FD mode's end-to-end SINR is given in (7). The distribution of ${{X_1}}$ is mentioned in \cite{krikidis2012full}
\begin{equation}
{F_{{X_1}}}(x) = 1 - \frac{1}{{1 + \eta x}}{e^{ - \frac{1}{{{P_S}{\lambda _1}}}x}}.
\end{equation}
%where $\eta  = \frac{{{\lambda _R}{P_R}}}{{{\lambda _1}{P_S}}}$.
%for the i.i.d.case, ${\eta _1}{\rm{ = }}{\eta _2} = \eta $, ${\lambda _1} = {\lambda _R}/{\eta _1} = {\lambda _R}/{\eta _2} = {\lambda _2} = \lambda $, and all channels are independent from each other.

The CDF of the end-to-end SINR is expressed as
\begin{equation}\label{fxx}
\Pr ({\gamma _F} > x) = \Pr (({X_1} - x)({P_R}{\gamma _2} - x) > {x^2} + x) = \frac{1}{{{\lambda _2}}}\int\limits_{x/{P_R}}^\infty  {\frac{{{e^{ - \frac{1}{{{P_S}{\lambda _1}}}(x + \frac{{{x^2} + x}}{{{P_R}{\gamma _2} - x}}) - \frac{1}{{{\lambda _2}}}{\gamma _2}}}}}{{1 + \eta (x + \frac{{{x^2} + x}}{{{P_R}{\gamma _2} - x}})}}} d{\gamma _2}.
\end{equation}

The integral in (\ref{fxx}) does not possess a closed-form solution in the scope of our knowledge.
The value of the integral is mainly decided by the exponent part, especially at high SNR.
We adopt Taylor's formula in \cite[eq.(1.112)]{zwillinger2014table} to derive the asymptotic result
\begin{equation}\label{appro1}
\frac{1}{{1 + \eta (x + \frac{{{x^2} + x}}{{{P_R}{\gamma _2} - x}})}}{\rm{ = }}\frac{1}{{1 + \eta x}}\frac{1}{{1 + \frac{\eta }{{1 + \eta x}}\frac{{{x^2} + x}}{{{P_R}{\gamma _2} - x}}}} \approx \frac{1}{{1 + \eta x}}(1 - \frac{\eta }{{1 + \eta x}}\frac{{{x^2} + x}}{{{P_R}{\gamma _2} - x}}).
\end{equation}

The integral (\ref{fxx}) is further obtained as
\begin{equation}
\Pr ({\gamma _F} > x) \approx \frac{{{e^{ - Cx}}}}{{{\lambda _2}(1 + \eta x)}}\int\limits_0^\infty  {{e^{ - (\frac{m}{{{\lambda _2}}} + \frac{{{x^2} + x}}{{{P_S}{P_R}{\lambda _1}}}\frac{1}{m})}}dm}  - \frac{{\eta ({x^2} + x){e^{ - Cx}}}}{{{\lambda _2}{P_R}{{(1 + \eta x)}^2}}}\int\limits_0^\infty  {\frac{1}{m}} {e^{^{ - (\frac{m}{{{\lambda _2}}} + \frac{{{x^2} + x}}{{{P_S}{P_R}{\lambda _1}}}\frac{1}{m})}}}dm,
\end{equation}
where $m = {\gamma _2} - \frac{x}{{{P_R}}}$ , $C = \frac{1}{{{\lambda _1}{P_S}}} + \frac{1}{{{\lambda _2}{P_R}}}$, with \cite[eq.(3.471.9)]{zwillinger2014table}, (\ref{fdcdf}) is obtained. Therefore, Lemma \ref{lemmafdcdf} can be obtained.

\section{Proof of Lemma \ref{lemmahybridcdf}}
We can write the CDF expressions of FD mode and HD mode as
\begin{equation}
\Pr ({\gamma _F} > x) = \int\limits_{x/{P_R}}^\infty  {\int\limits_{h({\gamma _2})}^\infty  {{f_{{X_1}}}({X_1}|{\gamma _2}){f_{{\gamma _2}}}({\gamma _2})d{X_1}d{\gamma _2}} },
\end{equation}

\begin{equation}
\Pr ({\gamma _H} > {x^2 + 2x}) = \int\limits_{({{x^2 + 2x})/{P_R}}}^\infty  {\int\limits_{g({\gamma _2})/{P_S}}^\infty  {{f_{{\gamma _1}}}({\gamma _1}|{\gamma _2}){f_{{\gamma _2}}}({\gamma _2})d{\gamma _1}d{\gamma _2}} },
\end{equation}
where $h(\gamma ) = x + \frac{{x + {x^2}}}{{{P_R}\gamma  - x}}$ , $g(\gamma ) = {x^2} + 2x + \frac{{{{({x^2} + 2x)}^2} + {x^2} + 2x}}{{{P_R}\gamma  - {x^2} - 2x}}$

The expression $\{ {\gamma _F} > x,{\gamma _H} > {x^2 + 2x}\} $ can be transformed into $\{ {\gamma _2} > \frac{x}{{{P_R}}},{\gamma _2} > \frac{{{x^2} + 2x}}{{{P_R}}},{\gamma _1} > \frac{{{P_R}{\gamma _R} + 1}}{{{P_S}}}h({\gamma _2}),{\gamma _1} > \frac{1}{{{P_S}}}g({\gamma _2})\} $.
As the value of $\gamma_F, {\gamma _H}$ are positive definite, we only consider the case when $x > 0$.
Therefore, $\{ {\gamma _F} > x,{\gamma _H} > {x^2 + 2x}\} $ can be further simplified as $\{ {\gamma _2} > \frac{{{x^2} + 2x}}{{{P_R}}},{\gamma _1} > \frac{{{P_R}{\gamma _R} + 1}}{{{P_S}}}h({\gamma _2}),{\gamma _1} > \frac{1}{{{P_S}}}g({\gamma _2})\} $
%derived after dividing it into two cases.
%As ${\gamma _F}, {\gamma _H}$ are positive, we only discuss the case

%\paragraph{$x > 1$}
%expression $\{ {\gamma _F} > x,{\gamma _H} > {x^2} + 2x\} $  equals $\{ {\gamma _2} > \frac{{{x^2}}}{{{P_R}}},{\gamma _1} > \frac{{{P_R}{\gamma _R} + 1}}{{{P_S}}}{h}({\gamma _2}),{\gamma _1} > \frac{1}{{{P_S}}}g({\gamma _2})\} $,
We define
\begin{equation}
A = \frac{1}{{{P_S}}}\left[ {({P_R}{\gamma _R} + 1){h}({\gamma _2}) - g({\gamma _2})} \right],
\end{equation}
when $A > 0$ , ${\gamma _R} > \frac{{x{\gamma _2} + {\gamma _2}}}{{{P_R}{\gamma _2} - {x^2} - 2x}}$ ,when $A < 0$ , $0< {\gamma _R} < \frac{{x{\gamma _2} + {\gamma _2}}}{{{P_R}{\gamma _2} - {x^2} - 2x}}$.

The distribution of $\{ {\gamma _F} > x,{\gamma _H} > {x^2 + 2x}\} $ splits into two sub-probabilities, $\{ {\gamma _2} > \frac{{{x^2} + 2x}}{{{P_R}}},{\gamma _1} > \frac{{{P_R}{\gamma _R} + 1}}{{{P_S}}}h({\gamma _2}),{\gamma _R} > \frac{{x{\gamma _2} + {\gamma _2}}}{{{P_R}{\gamma _2} - {x^2} - 2x}}\}  $ and $\{ {\gamma _2} > \frac{{{x^2} + 2x}}{{{P_R}}},{\gamma _1} > \frac{1}{{{P_S}}}g({\gamma _2}),0 < {\gamma _R} < \frac{{x{\gamma _2} + {\gamma _2}}}{{{P_R}{\gamma _2} - {x^2} - 2x}}\}  $ , denoted as ${I_1}$, ${I_2}$.

%\paragraph{$x <1$}
%as ${\gamma _R} > 0 > \frac{{x{\gamma _2} - {\gamma _2}}}{{{P_R}{\gamma _2} - {x^2}}}$ , the probability of $\{ {\gamma _F} > x,{\gamma _H} > {x^2}\} $  equals $\{ {\gamma _2} > \frac{x}{{{P_R}}},{\gamma _1} > \frac{{{P_R}{\gamma _R} + 1}}{{{P_S}}}{h}({\gamma _2}),{\gamma _R} > 0\} $ , denoted as ${I_3}$.

Consider ${I_1}$, we can write
\setlength{\arraycolsep}{0.0em}
\begin{eqnarray}
{I_1} && = \int\limits_{\frac{{{x^2} + 2x}}{{{P_R}}}}^\infty  {{f_{{\gamma _2}}}({\gamma _2})\int\limits_{\frac{{x{\gamma _2} + {\gamma _2}}}{{{P_R}{\gamma _2} - {x^2} - 2x}}}^\infty  {{f_{{\gamma _R}}}({\gamma _R})} } \int\limits_{\frac{{{P_R}{\gamma _R} + 1}}{{{P_S}}}{\rm{h}}({\gamma _2})}^\infty  {{f_{{\gamma _1}}}({\gamma _1})d{\gamma _1}d{\gamma _R}} d{\gamma _2}\nonumber\\
&& = \int\limits_{\frac{{{x^2} + 2x}}{{{P_R}}}}^\infty  {\frac{1}{{{\lambda _2}}}\frac{1}{{1 + \eta h({\gamma _2})}}{e^{ - (\frac{{{P_R}}}{{{P_S}{\lambda _1}}}h({\gamma _2}) + \frac{1}{{{\lambda _R}}})\frac{{x{\gamma _2} + {\gamma _2}}}{{{P_R}{\gamma _2} - {x^2} - 2x}} - \frac{1}{{{P_S}{\lambda _1}}}h({\gamma _2}) - \frac{1}{{{\lambda _2}}}{\gamma _2}}}d{\gamma _2}} ,
\end{eqnarray}
using the approximation in (\ref{appro1}), and $\frac{1}{{{\gamma _2} - x/{P_R}}} \approx \frac{1}{{{\gamma _2} - ({x^2 + 2x})/{P_R}}}$ in high SNR region, with the help of \cite[eq.(3.324.1)]{zwillinger2014table} and \cite[eq.(3.462.20)]{zwillinger2014table}, (\ref{I1}) is obtained.

Consider $I_2$, we can write
\begin{equation}
{I_2} = \int\limits_{\frac{{{x^2} + 2x}}{{{P_R}}}}^\infty  {{f_{{\gamma _2}}}({\gamma _2})\int\limits_0^{\frac{{x{\gamma _2} + {\gamma _2}}}{{{P_R}{\gamma _2} - {x^2} - 2x}}} {{f_{{\gamma _R}}}({\gamma _R})} } \int\limits_{\frac{1}{{{P_S}}}g({\gamma _2})}^\infty  {{f_{{\gamma _1}}}({\gamma _1})d{\gamma _1}d{\gamma _R}} d{\gamma _2},
\end{equation}
after a few mathematical manipulations, (\ref{I2}) is derived. Therefore, Lemma \ref{lemmahybridcdf} is proved.

%Consider $I_3$, we can write
%\begin{equation}
%{I_3} = \int\limits_{\frac{x}{{{P_R}}}}^\infty  {{f_{{\gamma _2}}}({\gamma _2})\int\limits_0^\infty  {{f_{{\gamma _R}}}({\gamma _R})} } \int\limits_{\frac{{{P_R}{\gamma _R} + 1}}{{{P_S}}}{h}({\gamma _2})}^\infty  {{f_{{\gamma _1}}}({\gamma _1})d{\gamma _1}} d{\gamma _R}d{\gamma _2},
%\end{equation}
%with the similar derivation as (\ref{fdcdf}), (\ref{I3}) is obtained.

\section{Proof of Proposition \ref{serpro}}
After substituting (\ref{cdf}) into (\ref{ser}) and adopting the approximation in the high SNR region that ${K_1}(z)$ converges to $\frac{1}{z}$, and that the value of ${K_0}(z)$ is comparatively small \cite[eq.(10.30)]{Frank}, which can be ignored for asymptotic analysis. (\ref{ser}) can be simplified as
\setlength{\arraycolsep}{0.0em}
\begin{eqnarray}\label{serpart}
\overline {SER} && = \frac{{{a_1}\sqrt {{a_2}} }}{{2\sqrt \pi  }}(\int\limits_0^\infty  {\frac{{{e^{ - {a_2}x}}}}{{\sqrt x }}dx}  - \int\limits_0^\infty  {\frac{{{e^{ - {a_2}x - Cx}}}}{{\sqrt x (1 + \eta x)}}dx}  - \int\limits_0^\infty  {\frac{{\eta x \cdot {e^{ - {a_2}x - C({x^2} + 2x) - \frac{1}{{{\lambda _1}{P_S}\eta }}(x + 1)}}}}{{\sqrt x (1 + \eta x)}}dx} )   \nonumber\\
 && = \frac{{{a_1}\sqrt {{a_2}} }}{{2\sqrt \pi  }}({l_1} - {l_2} - {l_3} ).
\end{eqnarray}

With the help of \cite[eq.(3.381.4)]{zwillinger2014table}, ${l_1}$ can be denoted as
\begin{equation}\label{ser1}
{l_1} = \int\limits_0^\infty  {\frac{{{e^{ - {a_2}x}}}}{{\sqrt x }}dx = {a_2}^{ - \frac{1}{2}}\Gamma (\frac{1}{2})}.
\end{equation}
%where $\Gamma (\cdot)$ is the Gamma Function \cite{zwillinger2014table}.
%Denoting ${l_2} = \int\limits_0^1 {\frac{{{e^{ - {a_2}x - C{x^2}}}}}{{\sqrt x }}dx}  $ , using the approximation ${e^{ - C{x^2}}} \approx \sum\limits_{k = 0}^3 {\frac{1}{{k!}}{{( - C{x^2})}^k}} $ when $x$ is around zero, and with the help of \cite[eq.(3.381.1)]{zwillinger2014table} ,  ${l_2}$ is given as
%\begin{equation}\label{ser2}
%{l_2} \approx \sum\limits_{k = 0}^3 {\frac{1}{{k!}}} {( - C)^k}\int\limits_0^1 {{x^{2k - \frac{1}{2}}}{e^{ - {a_2}x}}dx}  = \sum\limits_{k = 0}^3 {\frac{1}{{k!}}} {( - C)^k}{a_2}^{ - 2k - \frac{1}{2}}\gamma (2k + \frac{1}{2},{a_2}).
%\end{equation}

Denoting ${l_2} = \int\limits_0^\infty  {\frac{{{e^{ - {a_2}x - Cx}}}}{{\sqrt x (1 + \eta x)}}dx} $,
with the help of \cite[eq.(3.383.10)]{zwillinger2014table},  ${l_2}$ is given as
%, as the high SNR part of the SER integral is small owing to the property of Q-function, we can use the approximation $\frac{1}{{1 + x}} \approx {e^{ - x}} + \frac{1}{2}{x^2}{e^{ - \frac{5}{3}x}}$ \cite{7065323} when $x$ is around zero. With the help of \cite[eq.(3.382.4)]{zwillinger2014table}, \cite[eq.(3.384.3)]{zwillinger2014table}, ${l_3}$ is given as
\begin{equation}\label{ser3}
{l_2} = \frac{1}{\eta }\int\limits_0^\infty  {\frac{{{e^{ - ({a_2} + C)x}}}}{{\sqrt x (\frac{1}{\eta } + x)}}dx}  = \frac{1}{{\sqrt \eta  }}{e^{\frac{1}{\eta }({a_2} + C)}}\Gamma (\frac{1}{2})\Gamma (\frac{1}{2},\frac{1}{\eta }({a_2} + C)).
 \end{equation}
%where $\alpha  = \frac{\eta }{{\eta  + 1}}$, ${\mu _1} = C + {a_2} + \alpha$ , ${\mu _2} = C + {a_2} + \frac{5}{3}\alpha $.%, $\Gamma (a,x)$ is the incomplete Gamma Function and ${W_{\lambda ,\mu }}(z)$ is the Whittaker function \cite{zwillinger2014table}.

Denoting ${l_3} = \int\limits_0^\infty  {\frac{{\eta x \cdot {e^{ - {a_2}x - C({x^2} + 2x) - \frac{1}{{{\lambda _1}{P_S}\eta }}(x + 1)}}}}{{\sqrt x (1 + \eta x)}}dx} $ , when the SNR is high and $x$ is around zero, approximation $\frac{1}{{1 + x}} \approx {e^{ - x}} + \frac{1}{2}{x^2}{e^{ - \frac{5}{3}x}}$ \cite{7065323} is used, with \cite[eq.(3.462.1)]{zwillinger2014table}, ${l_3} $ is given as
 \setlength{\arraycolsep}{0.0em}
\begin{eqnarray}\label{ser4}
 {l_3}  \approx&& \int\limits_0^\infty  {\eta \sqrt x \left[ {{e^{ - \eta x}} + \frac{1}{2}{{\left( {\eta x} \right)}^2}{e^{ - \frac{5}{3}\eta x}}} \right]{e^{ - (\frac{1}{{{\lambda _1}{P_S}\eta }} + {a_2} + 2C)x - C{x^2} - \frac{1}{{{\lambda _1}{P_S}\eta }}}}dx}
\nonumber\\
 =&& \eta {e^{ - \frac{1}{{{\lambda _1}{P_S}\eta }}{\rm{ + }}\frac{{{\mu _1}^2}}{{8C}}}}{\left( {2C} \right)^{ - \frac{3}{4}}}\Gamma (\frac{3}{2}){D_{ - \frac{3}{2}}}(\frac{{{\mu _1}}}{{\sqrt {2C} }})
+ \frac{1}{2}{\eta ^3}{e^{ - \frac{1}{{{\lambda _1}{P_S}\eta }}{\rm{ + }}\frac{{{\mu _2}^2}}{{8C}}}}{\left( {2C} \right)^{ - \frac{7}{4}}}\Gamma (\frac{7}{2}){D_{ - \frac{7}{2}}}(\frac{{{\mu _2}}}{{\sqrt {2C} }}),
 \end{eqnarray}
where ${\mu _1} = {a_2} + 2C + \frac{1}{{{\lambda _1}{P_S}\eta }} + \eta $ , ${\mu _2} = {a_2} + 2C + \frac{1}{{{\lambda _1}{P_S}\eta }} + \frac{5}{3}\eta $.%, $\Gamma (a,x)$ is the incomplete Gamma Function and ${W_{\lambda ,\mu }}(z)$ is the Whittaker function \cite{zwillinger2014table}.

Substituting (\ref{ser1}), (\ref{ser3}) and (\ref{ser4}) into (\ref{serpart}), (\ref{serfinal}) can be obtained.

\section{Proof of Lemma \ref{sumrateintegral}}
After a few simplifications, we can derive
\begin{equation}\label{lemmaall}
\int\limits_b^\infty  {\frac{{{e^{ - C{x^2}}}}}{{x + a}}} dx = \frac{1}{{2a}}\int\limits_{{b^2}}^\infty  {\frac{{{e^{ - Cx}}}}{{\sqrt x }}dx - } \frac{1}{{2a}}(\int\limits_{{b^2}}^\infty  {\frac{{\sqrt x }}{{x - {a^2}}}{e^{ - Cx}}dx}  - a\int\limits_{{b^2}}^\infty  {\frac{1}{{x - {a^2}}}} {e^{ - Cx}}dx),
\end{equation}
with the help of \cite[eq.(3.361)]{zwillinger2014table}, the value of the first part can be obtained as
\begin{equation}\label{lemmapart1}
\int\limits_{{b^2}}^\infty  {\frac{{{e^{ - Cx}}}}{{\sqrt x }}} dx = \sqrt {\frac{\pi }{C}} \left[ {1 - \Phi (C{b^2})} \right].
\end{equation}

For integral $\int\limits_{{b^2}}^\infty  {\frac{{\sqrt x }}{{x - {a^2}}}{e^{ - Cx}}dx}$, as ${a^2}/x < {a^2}/{b^2} < 1$, using Taylor's formula $\frac{1}{{x - {a^2}}} = \frac{1}{{{a^2}}}\sum\limits_{k = 1}^\infty  {{{(\frac{{{a^2}}}{x})}^k}}$, the second part of (\ref{lemmaall}) can be derived as
\setlength{\arraycolsep}{0.0em}
\begin{eqnarray}\label{lemmapart2}
\int\limits_{{b^2}}^\infty  {\frac{{\sqrt x }}{{x - {a^2}}}{e^{ - Cx}}dx} &&  = \sum\limits_{k = 1}^\infty  {\int\limits_{{b^2}}^\infty  {{a^{2k - 2}}{x^{1/2 - k}}{e^{ - Cx}}dx} }  = \sum\limits_{k = 1}^\infty  {{a^{2k - 2}}{C^{k - 3/2}}\int\limits_{C{b^2}}^\infty  {{x^{1/2 - k}}{e^{ - x}}dx} } \nonumber\\
 && = \sum\limits_{k = 1}^\infty  {{a^{2k - 2}}{C^{k - \frac{3}{2}}}{{(C{b^2})}^{\frac{1}{4} - \frac{k}{2}}}{e^{ - \frac{{C{b^2}}}{2}}}{W_{\frac{1}{4} - \frac{k}{2},\frac{3}{4} - \frac{k}{2}}}(C{b^2})} .
 \end{eqnarray}

 With formula  ${E_1}(x) = {e^{ - x}}\int\limits_0^\infty  {\frac{{{e^{ - t}}}}{{t + x}}dt} $, the third part of (\ref{lemmaall}) can be derived as
\begin{equation}\label{lemmapart3}
\int\limits_{{b^2}}^\infty  {\frac{1}{{x - {a^2}}}} {e^{ - Cx}}dx = {E_1}(C({b^2} - {a^2})){e^{ - C{a^2}}}.
\end{equation}

Substituting (\ref{lemmapart1}), (\ref{lemmapart2}) and (\ref{lemmapart3}) into (\ref{lemmaall}), Lemma \ref{sumrateintegral} can be proved.

\section{Proof of Lemma \ref{lemmaw3p}}
As the upper limit of the integral $\int\limits_\rho ^{\frac{1}{\eta } + 1} {\frac{{{e^{ - C{x^2}}}}}{{x + \frac{1}{\eta } - \rho }}} dx$ only relate to $\eta $, in the high SNR region when $C{(1 + \frac{1}{\eta })^2}$ converges to zero, the approximation ${e^{ - C{x^2}}} \approx \sum\limits_{k = 0}^{{N_2}} {\frac{1}{{k!}}{{( - C{x^2})}^k}} $ is used to obtain the approximate value
\setlength{\arraycolsep}{0.0em}
\begin{eqnarray}\label{w3part}
\int\limits_\rho ^{1 + \frac{1}{\eta }} {\frac{{{e^{ - C{x^2}}}}}{{x + \frac{1}{\eta } - \rho }}} dx{\rm{ }} &&= \int\limits_0^{1 + \frac{1}{\eta } - \rho } {\frac{{{e^{ - C{{(x + \rho )}^2}}}}}{{x + \frac{1}{\eta }}}} dx \approx {e^{ - C{\rho ^2}}}\int\limits_0^{1 + \frac{1}{\eta } - \rho } {\frac{{{e^{ - 2C\rho x}}}}{{x + \frac{1}{\eta }}}\sum\limits_{k = 0}^{{N_2}} {\frac{1}{{k!}}{{( - C{x^2})}^k}} } dx\nonumber\\
&&  = \int\limits_{\frac{1}{\eta }}^{1 + \frac{2}{\eta } - \rho } {\sum\limits_{k = 0}^{{N_2}} {\sum\limits_{l = 0}^{2k} {\frac{1}{{k!}}{{( - C)}^k}{{( - \frac{1}{\eta })}^{2k - l}}(\begin{array}{*{20}{c}}
{2k}\\
l
\end{array}){x^{l - 1}}{e^{ - C{\rho ^{\rm{2}}}{\rm{ + }}\frac{{{\rm{2C}}\rho }}{\eta } - 2C\rho x}}} } } dx,
 \end{eqnarray}
where ${\varepsilon _1} = 1 + \rho$, ${\varepsilon _2} = 1 + \frac{1}{\eta }$. As $\sum\limits_{k = 0}^{{N_2}} {\sum\limits_{l = 0}^{2k} {( \cdot } ) = } \sum\limits_{k = 1}^{{N_2}} {\sum\limits_{l = 1}^{2k} {( \cdot } ) + } \sum\limits_{k = 0}^{{N_2}} {\sum\limits_{l = 0}^0 {( \cdot } )}  $ , (\ref{w3part}) can be divided into two parts. With the help of ${E_1}(x) = {e^{ - x}}\int\limits_0^\infty  {\frac{{{e^{ - t}}}}{{t + x}}dt} $ and \cite[eq.(3.381.1)]{zwillinger2014table} , (\ref{w3p}) is derived. Therefore, Lemma \ref{lemmaw3p} is proved.

\section{Proof of Proposition \ref{sumrate}}
After substituting (\ref{cdf}) into (\ref{sumrate-origin}) and with the help of ${K_1}(z)$ converges to $\frac{1}{z}$ when $z$ is around zero, ${\bar R}$ can be described as
\setlength{\arraycolsep}{0.0em}
\begin{eqnarray}\label{sumrate-part}
\bar R &&= \frac{1}{{\ln 2}}(\int\limits_0^\infty  {\frac{{{e^{ - Cx}}}}{{(1 + x)(1 + \eta x)}}dx}  + \int\limits_0^\infty  {\frac{{\eta x \cdot {e^{ - {\beta _4}}}}}{{(1 + x)(1 + \eta x)}}dx}  \nonumber\\
 &&   - \int\limits_0^\infty  {\frac{{2\eta ({x^2} + x){e^{ - Cx}}}}{{{\lambda _2}{P_R}(1 + x){{(1 + \eta x)}^2}}}{K_0}({\beta _1})dx}  + \int\limits_0^\infty  {\frac{{2\eta ({x^2} + x){e^{ - {\beta _4}}}}}{{{\lambda _2}{P_R}(1 + x){{(1 + \eta x)}^2}}}{K_0}({\beta _3})dx} )\nonumber\\
  && = \frac{1}{{\ln 2}}({w_1} + {w_2} - {w_3} + {w_4}).
\end{eqnarray}

%Denoting ${w_{\rm{1}}} = \frac{1}{{\ln 2}}(\int\limits_0^1 {\frac{{{e^{ - C{x^2}}}}}{{1 + x}}dx} $, using the first ${{N_1}}$ items of the Taylor expansion of ${e^{ - {x^2}}}$ when $x$ is around zero, ${w_1}$ can be obtained as
%\begin{equation}\label{w1}
%{w_{\rm{1}}} \approx \sum\limits_{k = 0}^{{N_1}} {\frac{1}{{k!}}{{( - C)}^k}} \ln 2 + \sum\limits_{k = 1}^{{N_1}} {\sum\limits_{i = 1}^{2k} {\frac{{{{\left( { - 1} \right)}^{3k - i}}}}{{i(k!)}}{C^k}(\begin{array}{*{20}{c}}
%{2k}\\
%i
%\end{array})} } ({2^i} - 1),
%\end{equation}
%we use ${{N_1}=3}$ in our analysis for approximation.

Denoting ${w_1} = \int\limits_0^\infty  {\frac{{{e^{ - Cx}}}}{{(1 + x)(1 + \eta x)}}dx}  $ , with the help of integral ${E_1}(x) = {e^{ - x}}\int\limits_0^\infty  {\frac{{{e^{ - t}}}}{{t + x}}dt} $ , ${w_1}$ can be derived as
\begin{equation}\label{w2}
{w_1} = \frac{{\rm{1}}}{{1 - \eta }}\left[ {{e^C}{E_1}(C) - {e^{\frac{C}{\eta }}}{E_1}(\frac{C}{\eta })} \right].
\end{equation}
%where ${E_1}( \cdot )$ is the exponential integral function \cite{zwillinger2014table}.

Denoting ${w_2} = \int_0^\infty  {\frac{{\eta x}}{{(1 + x)(1 + \eta x)}}{e^{ - C{x^2} - 2Cx - \frac{{x + 1}}{{{\lambda _1}{P_S}\eta }}}}dx}  $ , after a few mathematical simplifications, ${w_2}$ is given as
\setlength{\arraycolsep}{0.0em}
\begin{eqnarray}\label{w3}
{w_2} &&= \frac{{\eta {e^{ - \frac{1}{{{\lambda _1}{P_S}\eta }} + C{\rho ^2}}}}}{{\eta  - 1}}(\int\limits_\rho ^\infty  {\frac{{{e^{ - C{x^2}}}}}{{x + 1 - \rho }}} dx - \frac{1}{\eta }\int\limits_\rho ^\infty  {\frac{{{e^{ - C{x^2}}}}}{{x + \frac{1}{\eta } - \rho }}} dx)\nonumber\\
 &&  = \frac{{\eta {e^{C{\rho ^2} - \frac{1}{{{\lambda _1}{P_S}\eta }}}}}}{{\eta  - 1}}(\int\limits_\rho ^\infty  {\frac{{{e^{ - C{x^2}}}}}{{x + 1 - \rho }}} dx - \frac{1}{\eta }\int\limits_\rho ^{\frac{1}{\eta } + 1} {\frac{{{e^{ - C{x^2}}}}}{{x + \frac{1}{\eta } - \rho }}} dx - \frac{1}{\eta }\int\limits_{\frac{1}{\eta } + 1}^\infty  {\frac{{{e^{ - C{x^2}}}}}{{x + \frac{1}{\eta } - \rho }}} dx),
\end{eqnarray}
where $\rho  = 1 + \frac{1}{{2C{\lambda _1}{P_1}\eta }},1 < \rho  < 1 + \frac{1}{{2\eta }}$,
$|\frac{1}{\eta } - \rho | < \max \{ |\frac{1}{\eta } - 1|,|\frac{1}{{2\eta }} - 1|\}  < |\frac{1}{\eta } + 1|$.
The approximate value of $\int\limits_\rho ^{\frac{1}{\eta } + 1} {\frac{{{e^{ - C{x^2}}}}}{{x + \frac{1}{\eta } - \rho }}} dx$ is given in Lemma \ref{lemmaw3p} , we use ${N_2}=6$ for approximation.
The exact expression of $\int\limits_\rho ^\infty  {\frac{{{e^{ - C{x^2}}}}}{{x + 1 - \rho }}} dx $ , $\int\limits_{\frac{1}{\eta } + 1}^\infty  {\frac{{{e^{ - C{x^2}}}}}{{x + \frac{1}{\eta } - \rho }}} $ can be derived using Lemma \ref{sumrateintegral}, we use the first ${{N_1}}$ , ${{N_3}}$ items to derive the approximate value.
When $\eta  = 0.2$ , we use ${N_1}=3,{N_3} = 6$ for approximation.
%For the special case $1 - \rho = 0$ , $\int_{1 + \rho}^\infty  {\frac{{{e^{ - C{x^2}}}}}{{x + 1 -\rho}}dx}  = \frac{1}{2}{E_1}(4C)$.
Therefore, the value of integral ${w_2}$ is obtained.

Denoting ${w_3} = \int\limits_0^\infty  {\frac{{2\eta ({x^2} + x){e^{ - Cx}}}}{{{\lambda _2}{P_R}(1 + x){{(1 + \eta x)}^2}}}{K_0}({\beta _1})dx} $, after adopting $m = \frac{2}{{\sqrt {{\lambda _1}{\lambda _2}{P_S}{P_R}} }}x = {C_2}x$, we can derive
\begin{equation}\label{w4}
 {w_3} = \int\limits_0^\infty  {\frac{{2\eta ({x^2} + x){e^{ - Cx}}}}{{{\lambda _2}{P_R}(1 + x){{(1 + \eta x)}^2}}}{K_0}({\beta _1})dx}  = \int\limits_0^\infty  {\frac{{2\eta m{e^{ - mC/{C_2}}}}}{{{\lambda _2}{P_R}{C_2}^2{{(1 + \eta m/{C_2})}^2}}}} {K_0}(\sqrt {m(m + {C_2})} )dm,
\end{equation}
with the help of \cite[eq.(6.647.1)]{zwillinger2014table}, ${w_3}$ can be obtained.
%\begin{equation}\label{w3}
%{w_3} \approx \frac{2}{{{\lambda _2}{P_R}{C_2}}}{e^{\frac{C}{{2\eta }}}}{\Gamma ^2}(2){W_{ - \frac{3}{2},0}}({z_1}){W_{ - \frac{3}{2},0}}({z_2}).
%\end{equation}

For ${w_4} = \int\limits_0^\infty  {\frac{{2\eta ({x^2} + x){e^{ - {\beta _4}}}}}{{{\lambda _2}{P_R}(1 + x){{(1 + \eta x)}^2}}}{K_0}({\beta _3})dx} $, as in the high SNR region, ${K_0}({\beta _3})$ converges to zero, ${w_4}$ is comparatively small compared with other parts in (\ref{sumrate-part}) and can be ignored in our derivation.

Substituting (\ref{w2}), (\ref{w3}), (\ref{w4})  into (\ref{sumrate-part}), Proposition \ref{sumrate} is derived.

\end{document}